\documentclass[prd,twocolumn,nofootinbib,showpacs]{revtex4} 

\usepackage{graphicx,amsmath,amsfonts,amssymb,revsymb,dcolumn,epsfig,bm}

\begin{document}

\title{Searching for non-Gaussianity in the WMAP data}
\author{A. Bernui}\email{bernui@das.inpe.br}
\affiliation{Centro Brasileiro de Pesquisas F\'{\i}sicas\\
Rua Dr.\ Xavier Sigaud 150 \\
22290-180 Rio de Janeiro -- RJ, Brazil}

\author{M.J. Rebou\c{c}as}\email{reboucas@cbpf.br}
\affiliation{Centro Brasileiro de Pesquisas F\'{\i}sicas\\
Rua Dr.\ Xavier Sigaud 150 \\
22290-180 Rio de Janeiro -- RJ, Brazil}

\date{\today}

\begin{abstract}
Some analyses of recent cosmic microwave background (CMB) data have provided
hints that there are deviations from Gaussianity in the WMAP CMB temperature
fluctuations.
Given the far reaching consequences of such a non-Gaussianity for our
understanding of the physics of the early universe, it is important to employ
alternative indicators in order to determine whether the reported non-Gaussianity
is of cosmological origin, and/or extract further information that may be helpful
for identifying its causes.
We propose two new non-Gaussianity indicators, based on skewness and kurtosis of
large-angle patches of CMB maps, which provide a measure of departure from
Gaussianity on large angular scales.
A distinctive feature of these indicators is that they provide sky maps of
non-Gaussianity of the CMB temperature data, thus allowing a possible additional
window into their origins.
Using these indicators, we find no significant deviation from Gaussianity
in the three and five-year WMAP ILC map with \emph{KQ75} mask, while the
ILC unmasked map exhibit deviation from Gaussianity, quantifying
therefore the  WMAP team recommendation to employ the new mask \emph{KQ75}
for tests of Gaussianity.
We also use our indicators to test for Gaussianity the single frequency foreground
unremoved WMAP three and five-year maps, and show that the K and Ka maps
exhibit clear indication of deviation from Gaussianity even with
the \emph{KQ75} mask.
We show that our findings are robust with respect to the details of the
method.  
\end{abstract}

\pacs{98.80.Es, 98.70.Vc, 98.80.-k}

\maketitle

\section{Introduction}

Within the standard approach to cosmological modelling the suggestion that
the Universe underwent a brief period of rapid acceleration expansion%
~\cite{Inflation-1st-refs} before the epoch of primordial nucleosynthesis
has become an essential building block of the standard cosmological model.
Besides solving the so-called flatness, horizon and monopole problems,
such a period of cosmological inflation  provides a mechanism for the
production of the primordial density fluctuations, which seeded
the observed cosmic microwave background (CMB) anisotropies
and the formation of large-scale structure in the Universe.

There are more than one hundred inflationary models (see, e.g.,
the review articles Refs.~\cite{Inflation-reviews}), among which
the simple ones are based on a slowly-rolling single scalar field.
An important prediction of a number of these simple models is that they can
generate only tiny non-Gaussianity, which should be undetectable
in the Wilkinson Microwave Anisotropy Probe (WMAP) CMB data%
~\cite{Gauss_Single-field}.
There are, however, a large class of inflationary models that can generate
non-Gaussianity at a level detectable by WMAP~\cite{Non-standard-models}.
These scenarios comprise models based upon a wide range of mechanisms,
including special features of the inflation potential, multiple scalar fields,
non-canonical kinetic terms, and  non-adiabatic fluctuations 
(see the review Ref.~\cite{Bartolo2004} and references therein).
Thus, the detection of non-Gaussianity
in CMB data may potentially be useful to discriminate inflationary models
and shed light on the physics of the early universe.

In the statistical analyses by using one, three and five-year%
~\cite{wmap1,wmap3,wmap5,Gauss-consist} CMB measurements  along with some
different statistical tools, the WMAP team have found that the CMB data are
consistent with Gaussianity.
However, some recent analyses  have provided clear hints
that there are significant deviations from Gaussianity in the WMAP data.
Clearly the study of detectable non-Gaussianities in the WMAP data must
take into account that they may have non-cosmological origins as, for
example, unsubtracted contamination from galactic diffuse foreground
emission~\cite{Chiang-et-al2003,Naselsky-et-al2005}
and unconsidered point sources~\cite{Eriksen-et-al2004}.
If they turn out to have a cosmological origin, however, this could
have far-reaching consequences on our description of the Universe,
particularly on the inflationary picture.

In view of this, a great deal of effort has recently gone into verifying
the existence of such non-Gaussianity by employing several different
statistical signatures of non-Gaussianity in its various forms (see, e.g.,
Refs.~\cite{Some_non-Gauss-refs} and related Refs.~\cite{Non-Gauss_related}).
Apart from revealing the existence of non-Gaussianity in CMB data,
different statistical tools are sensitive to different systematics
and may be useful in determining their origins. In addition, different
indicators can in principle provide information about the multiple types
of non-Gaussianity that may be present in CMB data.
It is therefore important to test the data for deviations from
Gaussianity by using a range of different statistical tools to
identify any non-Gaussian signals on the CMB sky.

In this paper, we propose new large-angle non-Gaussianity indicators, based
on skewness and kurtosis of large-angle spherical-shaped patches of CMB maps,
which provide a measure of departure from Gaussianity on large angular scales.
A distinctive feature of these indicators is that they provide sky maps
(directional information) of non-Gaussianity of the CMB temperature fluctuations,
thus allowing a possible additional window into their causes.
Using these indicators, we find no significant deviation from Gaussianity
in the WMAP three and five-year ILC \emph{KQ75} masked map, but
ILC unmasked map exhibit deviation from Gaussianity.
On the other hand, our indicators reveal deviations from Gaussianity of
different degree in the three and five-year single frequency K and Ka
\emph{KQ75} masked maps, which is consistent with the fact that even these
masked maps are still foreground contaminated at some level.

The structure of the paper is as follows. In Sec.~\ref{Indicators}
we introduce our non-Gaussianity indicators. Section~\ref{Indic_vs_data}
contains the results of applying our statistical indicators to the
three-year and five-year WMAP data, and f\/inally in Sec.~\ref{Conclusions}
we present the summary of our main results and conclusions.

\section{Non-Gaussianity Indicators and Maps}
\label{Indicators}

In this section we construct two statistical indicators that measure
the large-angle deviation from Gaussianity of CMB temperature fluctuation
patterns.

The main underlying idea in the construction of our non-Gaussianity indicators
and the associated maps, is that the simplest ways of describing the deviation
from symmetry  about the CMB mean temperature, and a non-Gaussian degree of
peakness are by calculating, respectively,  the skewness $S=\mu_3/\sigma^3$,
and the kurtosis $K=\mu_4/\sigma^4\!\!-\!3$ from the data, where $\mu_3$ and $\mu_4$
are the third and fourth central moments of the temperature anisotropies
distribution, and $\sigma$ is the variance.
Clearly calculating $S$ and $K$ for the whole
celestial sphere would simply yield two dimensionless numbers
describing the asymmetry of the data about the CMB average temperature.

However, one can go a step further and obtain directional information
about deviation from Gaussianity if instead we take a discrete set of
points $\{j=1, \ldots ,N_\mathrm{c} \}$ homogeneously distributed on the
celestial sphere $S^2$ as the center of spherical caps of a given
aperture $\gamma$ and calculate $S_j$ and $K_j$ for
each cap. The values $S_j$ and $K_j$ can then be viewed as
measures of the non-Gaussianity in the direction $(\theta_j, \phi_j)$
of the center of the cap $j\,$.
Such study of the individual caps can thus provide information
($2 N_\text{c}$ numbers) about possible large-angle violation of
Gaussianity in the CMB data.
A more systematic study can be made by taking the above set of
points $\{j\}$ as the center of the pixels for a  homogeneous
pixelization of $S^2$, and by choosing caps of large-angle aperture
to scan the whole celestial sphere with steps equal to the separation
between the centers of adjacent pixels.%
\footnote{Note that here this pixelization is only a practical way
of choosing the centers of the caps homogeneously distributed on $S^2$.
It is not related to the pixelization of the CMB maps.}
In this way, we construct two scalar discrete functions $S$ and $K$
defined over the whole celestial sphere that encode measures of
non-Gaussianity in CMB data.

This constructive process can be formalized as follows.
Let $\Omega_j \equiv \Omega(\theta_j,\phi_j;\gamma) \in
S^2$ be a spherical cap, with an aperture of $\gamma$ degrees,
centered at $(\theta_j,\phi_j)$, for $j=1, \ldots, N_\text{c}$.
Define the scalar functions $S: \Omega_j \mapsto \mathbb{R} $
and $K: \Omega_j \mapsto \mathbb{R} $, that
assign to the $j^{\,\rm{th}}$ cap, centered at $(\theta_j,\phi_j)$, two
real numbers $S_j$ and $K_j$ given by
\begin{eqnarray}
S_j  & \equiv & \frac{1}{N_{\mbox{p}} \,\sigma^3_j } \sum_{i=1}^{N_{\mbox{\footnotesize p}}}
\left(\, T_i\, - \overline{T} \,\right)^3 \;,  \\
K_j  & \equiv & \frac{1}{N_{\mbox{p}} \,\sigma^4_j } \sum_{i=1}^{N_{\mbox{\footnotesize p}}}
\left(\,  T_i\, - \overline{T} \,\right)^4 - 3 \;,
\end{eqnarray}
where $N_{\mbox{p}}$ is the number of pixels in the $j^{\,\rm{th}}$ cap,
$T_i$ is the temperature at the $i^{\,\rm{th}}$ pixel, $\overline{T}$ is
the CMB mean temperature, and $\sigma_j$ is the standard deviation for
each $j$.

\begin{figure*}[htb!]
\begin{center}
\includegraphics[width=5.2cm,height=8.2cm,angle=90]{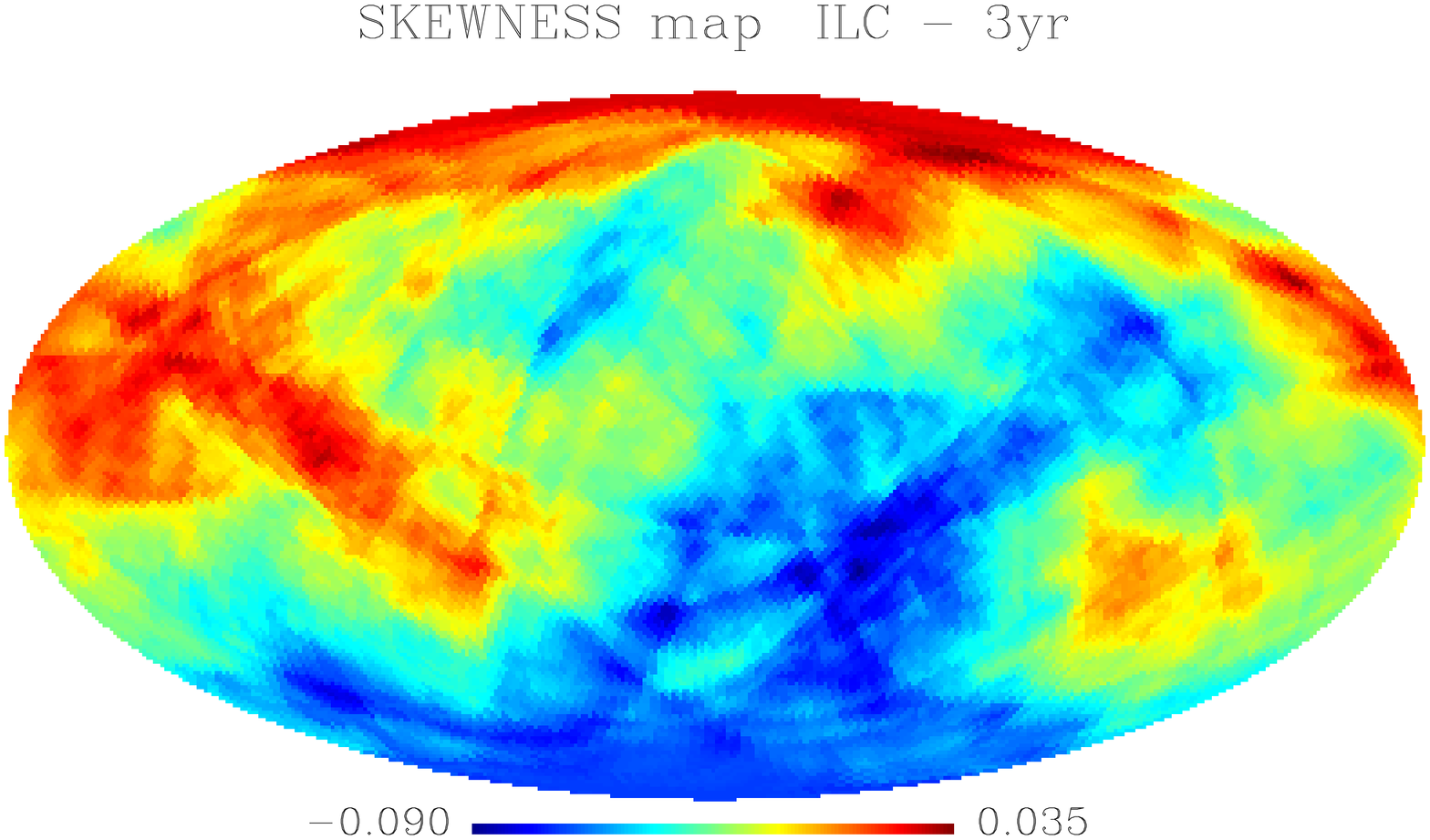}
\hspace{5mm}
\includegraphics[width=5.2cm,height=8.2cm,angle=90]{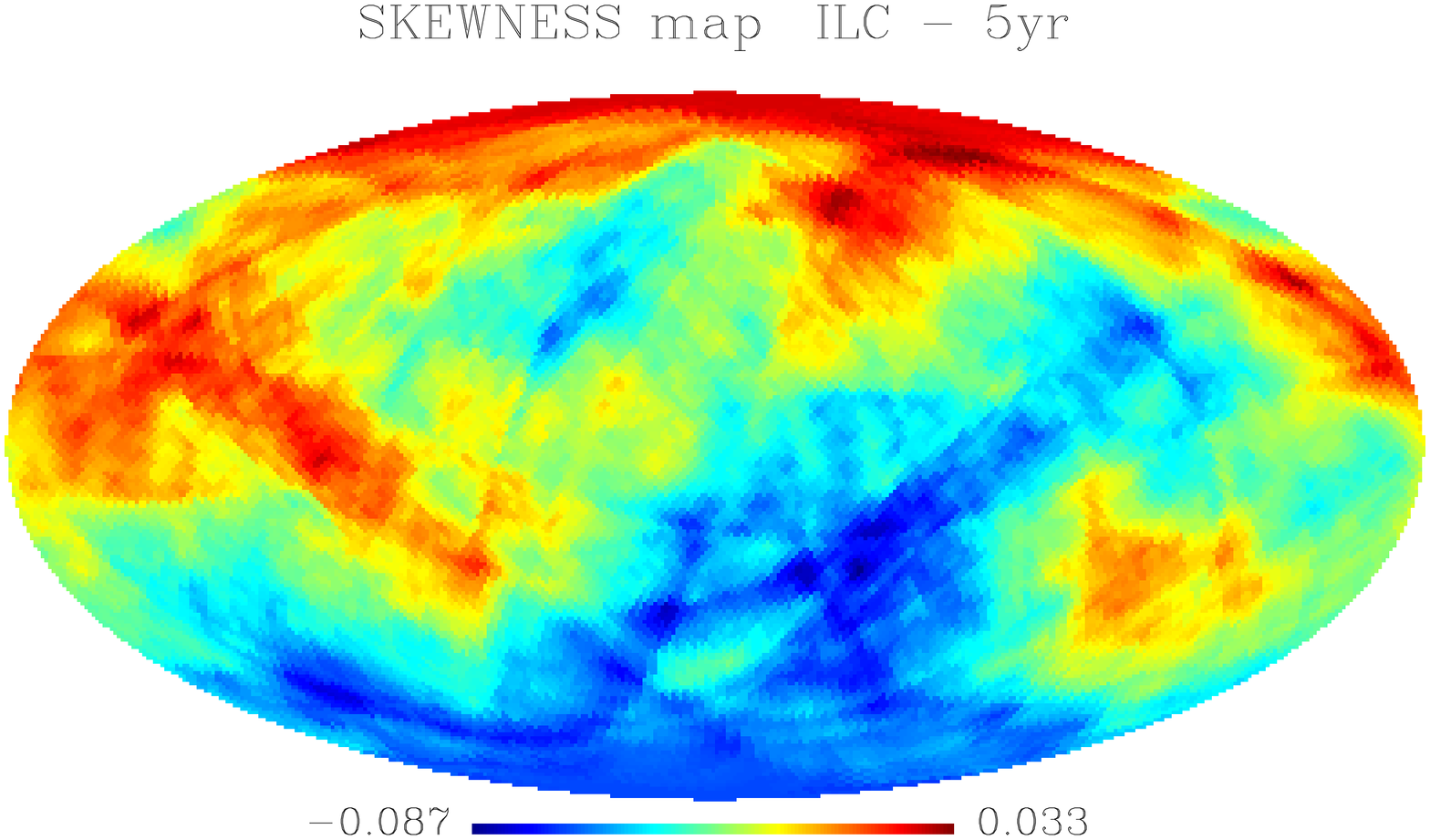}
\caption{\label{Fig1} Skewness indicator maps from
the WMAP three (left panel) and five-year (right panel) ILC maps
with mask \emph{KQ75}.}
\end{center}
\end{figure*}

\begin{figure*}[htb!]
\begin{center}
\includegraphics[width=5.2cm,height=8.2cm,angle=90]{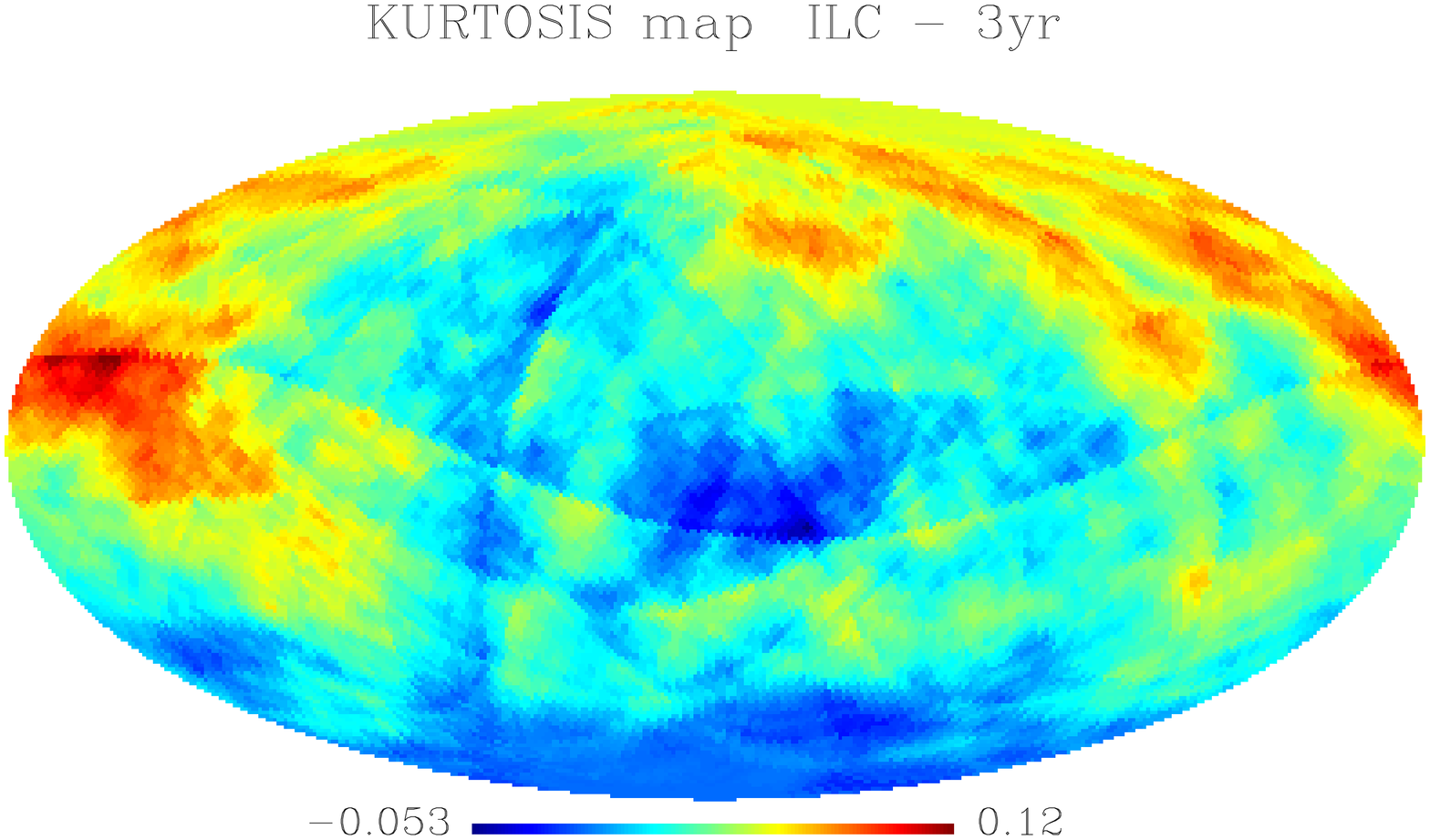}
\hspace{5mm}
\includegraphics[width=5.2cm,height=8.2cm,angle=90]{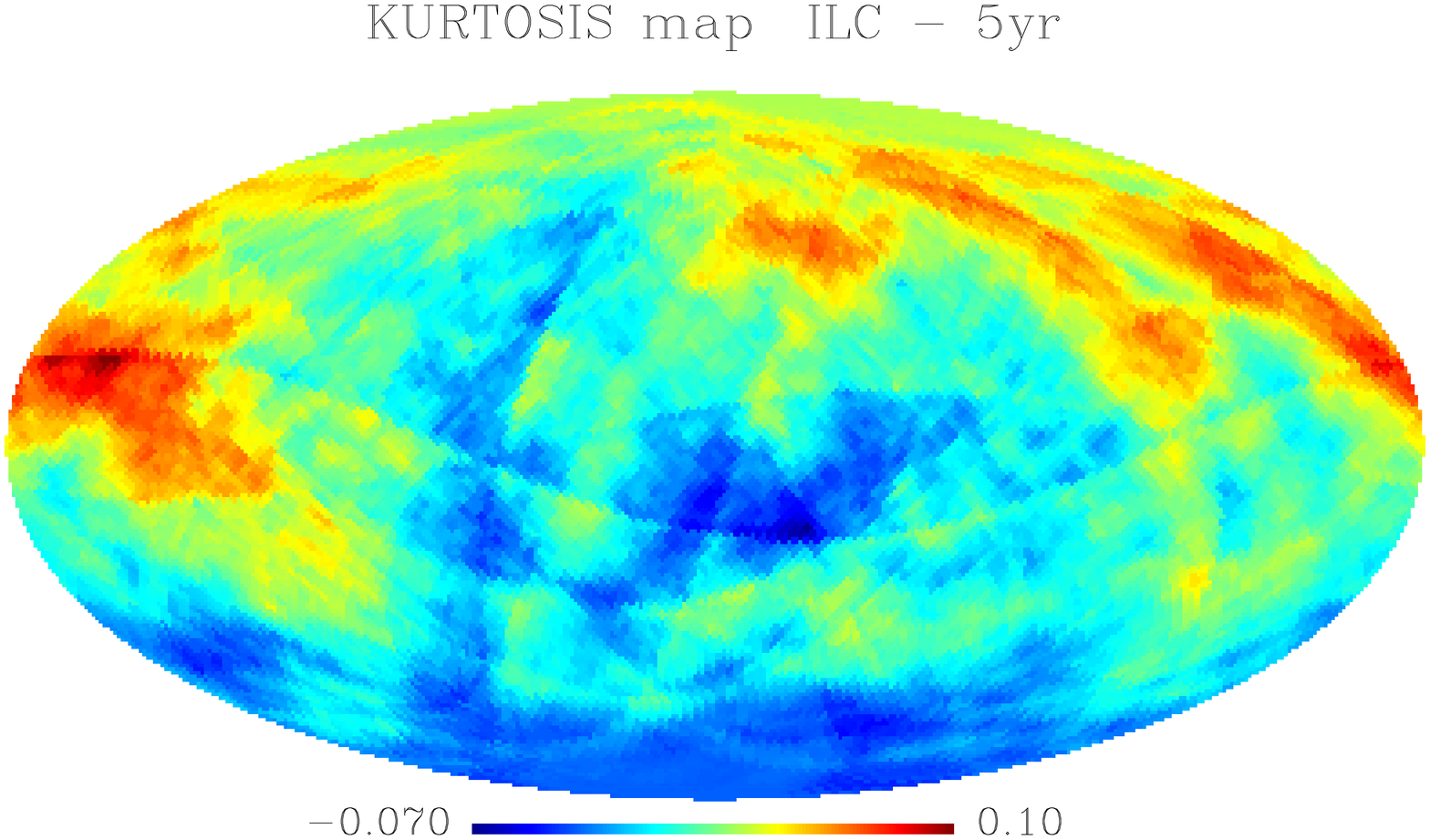}
\caption{\label{Fig2} Kurtosis indicator maps
from the WMAP three (left panel) and five-year (right panel) ILC
maps with mask \emph{KQ75}.}
\end{center}
\end{figure*}

We now use the above homogeneously distributed points on $S^2$ to
scan the celestial sphere with evenly distributed spherical caps
(of a chosen aperture $\gamma$)
to calculate $S_j$ and $K_j$ for each cap.
Clearly, the numbers $S_j$ and $K_j$ obtained in this way for each cap
can then be viewed as a measure of non-Gaussianity in the direction of
the center of that cap $(\theta_j, \phi_j)$.
Patching together the $S_j$ and $K_j$ values for each cap, we obtain
the indicators that are discrete functions $S = S(\theta,\phi)$ and
$K = K(\theta,\phi)$ defined over the celestial sphere, which
measure the deviation from Gaussianity as a function of direction
$(\theta,\phi)$. In this way, $S = S(\theta,\phi)$ and $K = K(\theta,\phi)$
give a scalar directional measure of non-Gaussianity over the celestial
sphere.

Now, since $S = S(\theta,\phi)$ and $K = K(\theta,\phi)$ are discrete
scalar functions defined on $S^2$ they can also be viewed as maps
of non-Gaussianity, and  we can expand each of these functions in
their spherical harmonics and calculate their angular power spectrum.
Thus, for the skewness function $S = S(\theta,\phi)$, for example,
one has
\begin{equation}
S (\theta,\phi) = \sum_{\ell=0}^\infty \sum_{m=-\ell}^{\ell}
b_{\ell m} \,Y_{\ell m} (\theta,\phi) \; ,
\end{equation}
and can calculate the corresponding angular power spectrum
\begin{equation}
S_{\ell} = \frac{1}{2\ell+1} \sum_m |b_{\ell m}|^2 \; ,
\end{equation}
in order to further quantify the angular scale information regarding
the deviation from Gaussianity of CMB data. Clearly, one can similarly expand
the kurtosis function $K = K(\theta,\phi)$ and calculate its angular power
spectrum $K_{\ell}$.
It then follows that, if a large-scale non-Gaussianity is present in the
original temperature distribution, it should significantly affect
the $S$ and $K$ maps on the corresponding angular scales.

In the next section we shall apply the indicators $S = S(\theta,\phi)$
and $K = K(\theta,\phi)$ to both WMAP three and five-year data.

\section{Non-Gaussianity Indicators and WMAP data}
\label{Indic_vs_data}

\begin{figure*}[hbt!]
\begin{center}  
\includegraphics[width=5.2cm,height=8.2cm,angle=90]{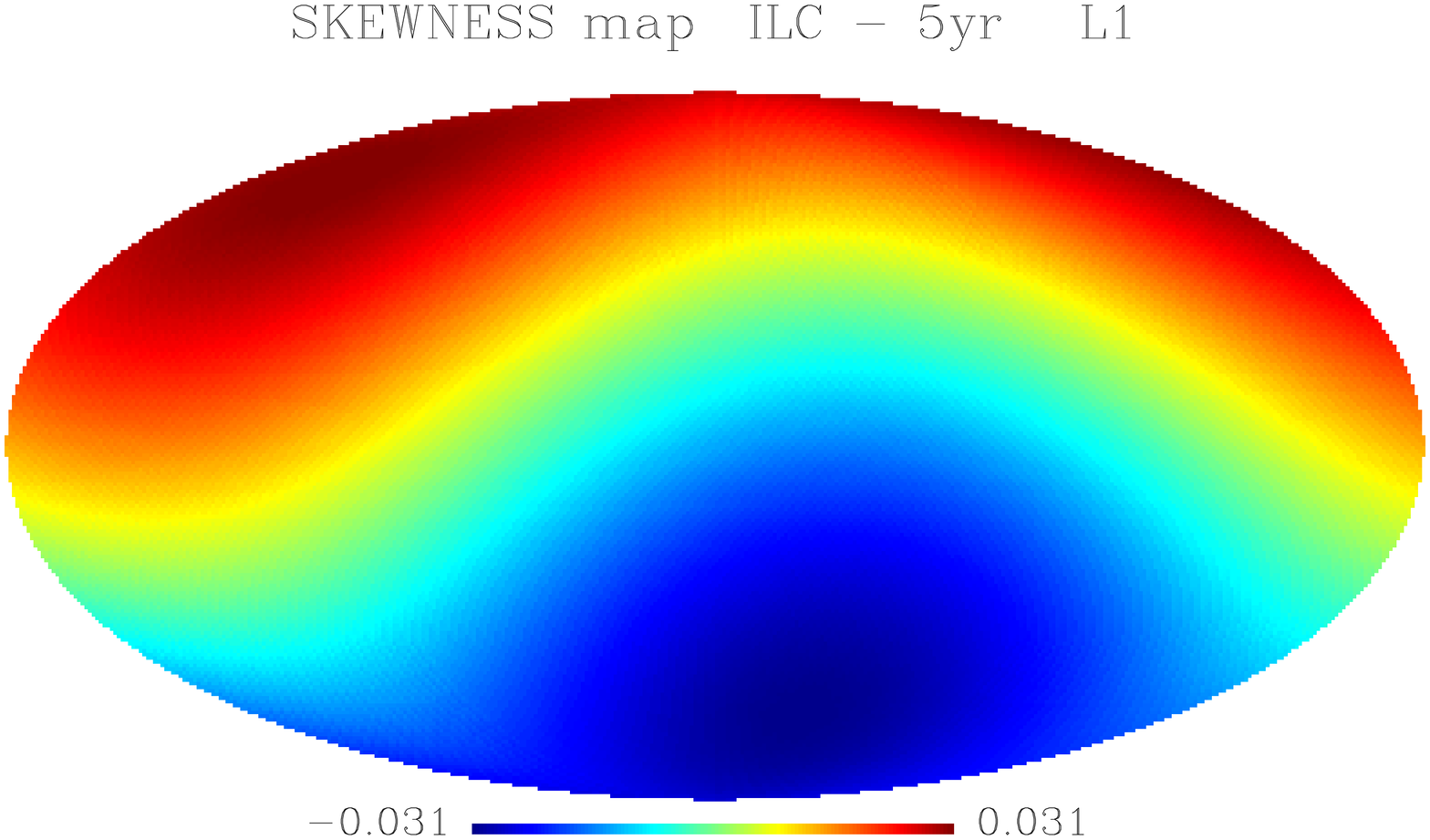}
\hspace{3mm}
\includegraphics[width=5.2cm,height=8.2cm,angle=90]{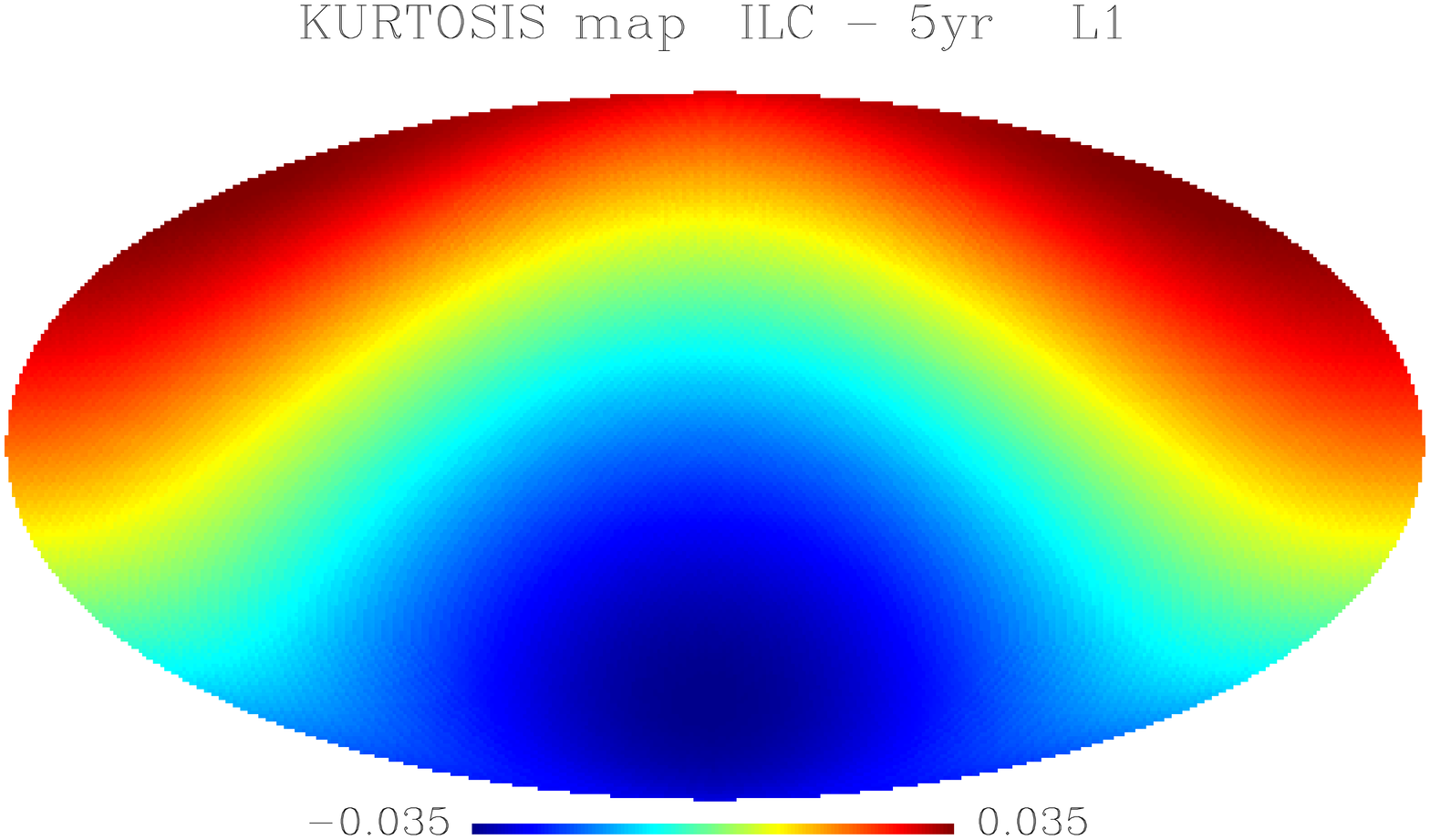}
\vskip 3mm
\includegraphics[width=5.2cm,height=8.2cm,angle=90]{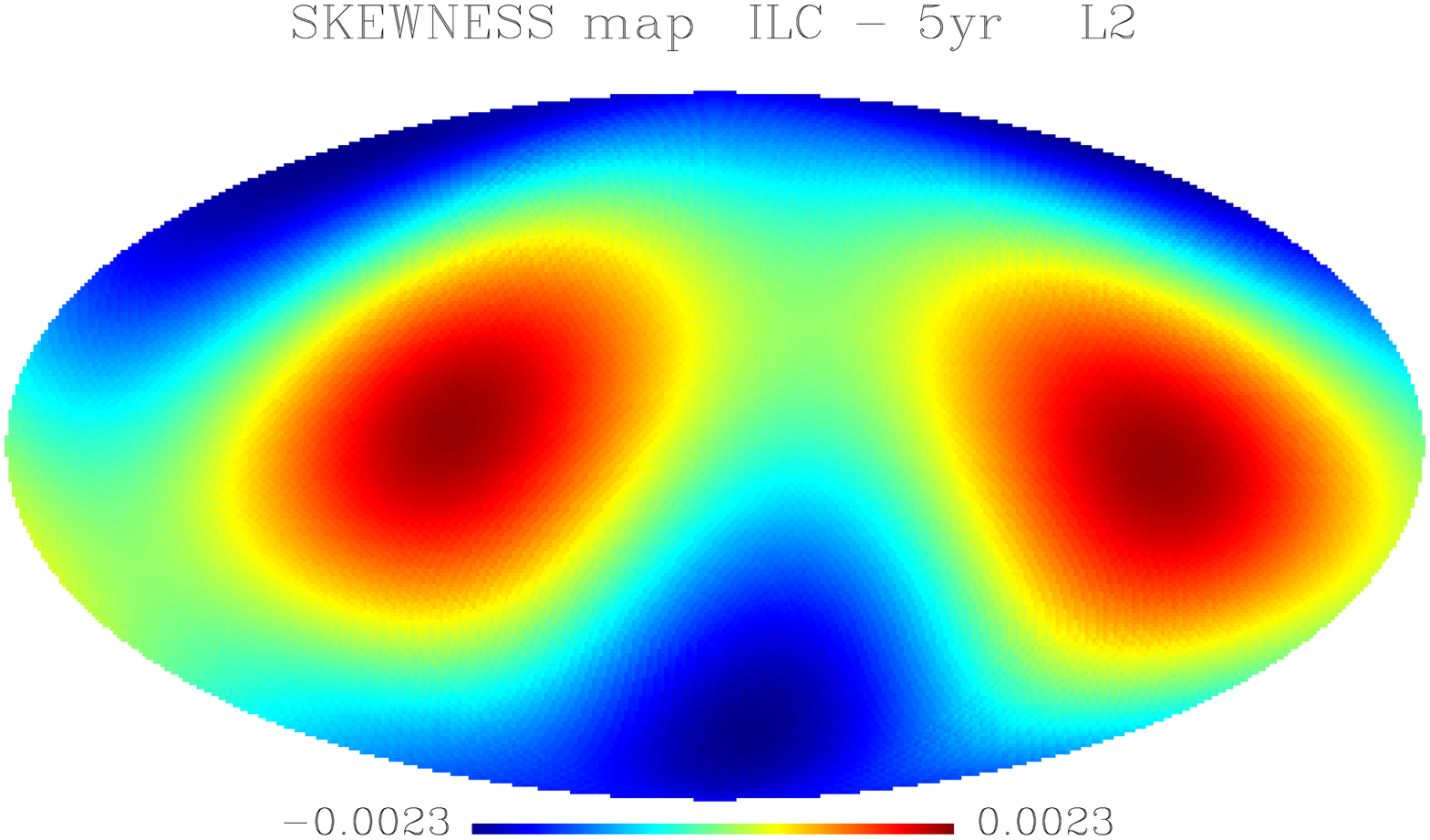}
\hspace{3mm}
\includegraphics[width=5.2cm,height=8.2cm,angle=90]{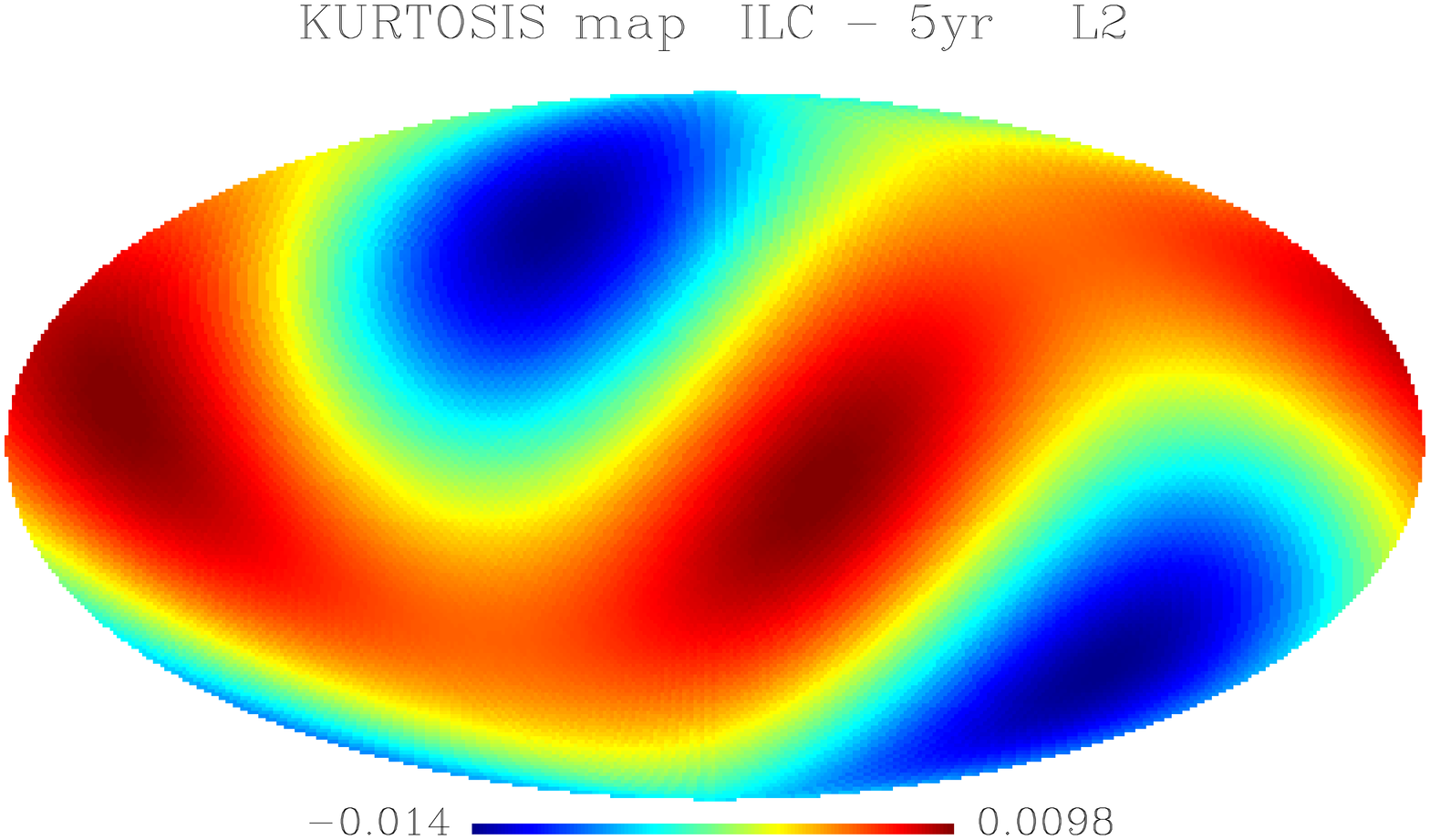}
\vskip 3mm
\includegraphics[width=5.2cm,height=8.2cm,angle=90]{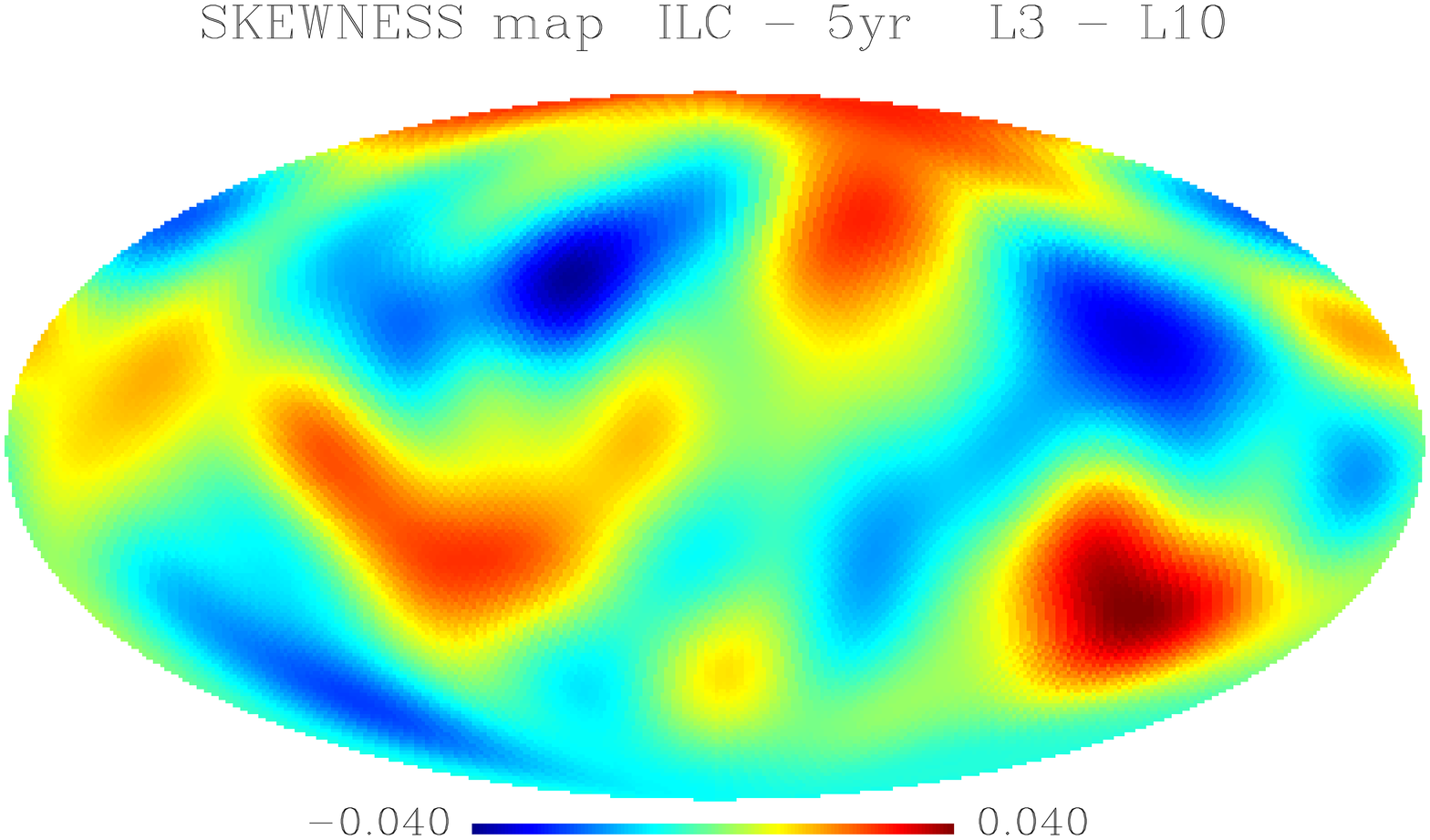}
\hspace{3mm}
\includegraphics[width=5.2cm,height=8.2cm,angle=90]{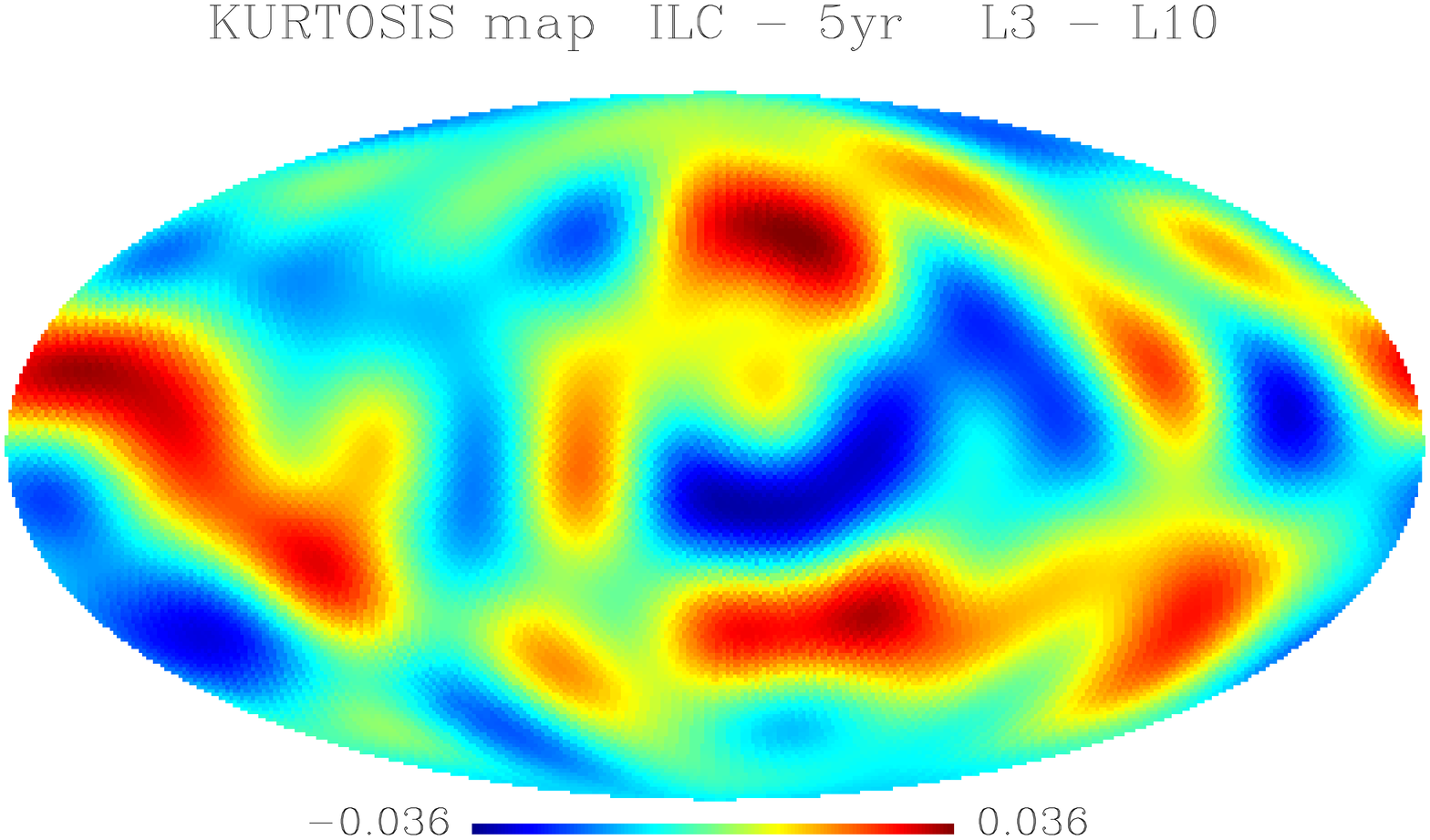}
\caption{\label{Fig3} Depicted the dipole, quadrupole and the remaining low $\ell$
components for the skewness (left panels) and kurtosis (right panels) maps
obtained from the five-year WMAP ILC with mask \emph{KQ75}.} 
\end{center}
\end{figure*}

The WMAP team have produced high angular resolution maps of the
CMB temperature fluctuations in five frequency bands: K--band
($22.8$ GHz), Ka--band ($33.0$ GHz), Q--band ($40.7$ GHz),
V--band ($60.8$ GHz), and W--band ($93.5$ GHz).
Following the WMAP team recommendation for tests of Gaussianity%
~\cite{wmap5,wmap5-Gold-et-al},
we used the Internal Linear Combination (ILC) maps of both the three-year
and the five-year CMB data~\cite{ILC-WMAP-refs} along with  the new mask
\emph{KQ75}, which is slightly more conservative than the \emph{Kp0}, i.e.,
\emph{KQ75} sky cuts is $28.4\%$ while \emph{Kp0} cuts is $24.5\%\,$.
We also test for non-Gaussianity the ILC full-sky five-year map along
with the five frequency foreground uncleaned five-year maps with and
without the \emph{KQ75} mask. 
In all cases we chose the HEALPix parameter $N_{\mbox{side}}=256$%
~\cite{Gorski-et-al-2005}, which corresponds to a partition of
the celestial sphere into $786\,432$ pixels.

In our calculations of skewness and kurtosis indicator maps
(hereafter referred to as $S-$map and $K-$map) from each
CMB map we have scanned the celestial sphere with spherical
caps of aperture  $\gamma = 90^{\circ}$, centered at
$N_\text{c}=768,\, 3\,072$ and $12\,288$ points  homogeneously
generated on the sphere by using  HEALPix.
However, to avoid repetition we only present a detailed analysis
for $N_\text{c}=12\,288$ in the following.

Figures~\ref{Fig1} and~\ref{Fig2} show, respectively, the Mollweide
projection of the $S-$map and $K-$map in galactic coordinates obtained
from the ILC WMAP three (left panels) and five-year (right panels)  maps
with the \emph{KQ75}  mask. They clearly show that the $S(\theta,\phi)$ and
$K(\theta,\phi)$ distributions of hot and cold spots (higher and lower
values) for the indicators are not evenly distributed in the celestial
sphere, suggesting at first sight non-Gaussianity of the ILC masked data.
The comparison between the two $S-$maps (Fig.~\ref{Fig1}) and 
the two $K-$maps (Fig.~\ref{Fig2}) shows a great number of
similarities for each pair of maps of the indicators, which is a very first
indication of the robustness of our results with respect to the three
and five-year WMAP data.%
\footnote{It is interesting to note the presence of great circles
of unknown origin near the galactic plane in the $K$ maps, and in a
less accentuated way in the $S$ maps.}
Figures~\ref{Fig1} and~\ref{Fig2}  are also suggestive of large-scale
components in the maps of both indicators.
\begin{figure*}[htb!]
\begin{center}
\includegraphics[width=8.8cm,height=5.6cm]{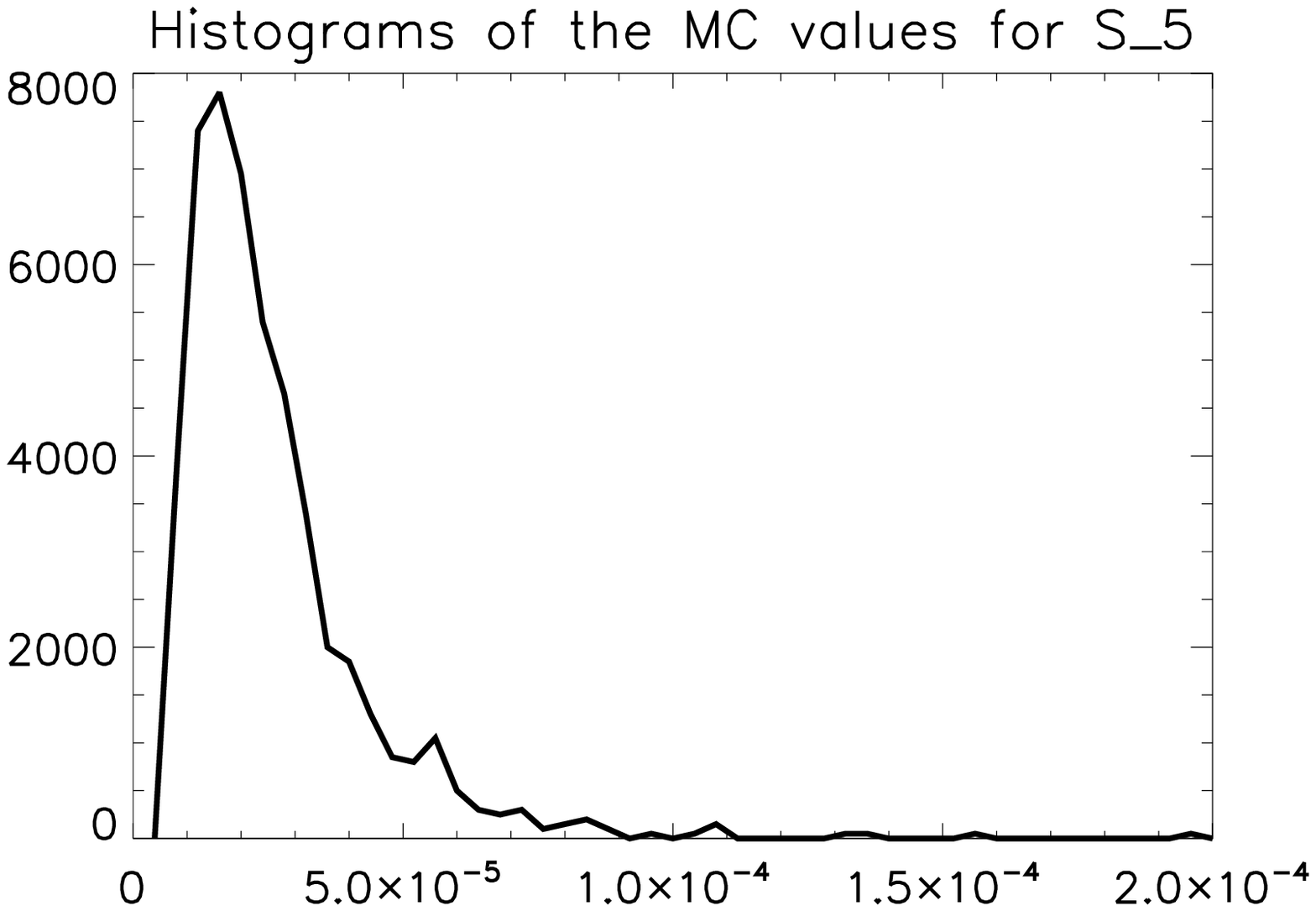}
\includegraphics[width=8.8cm,height=5.6cm]{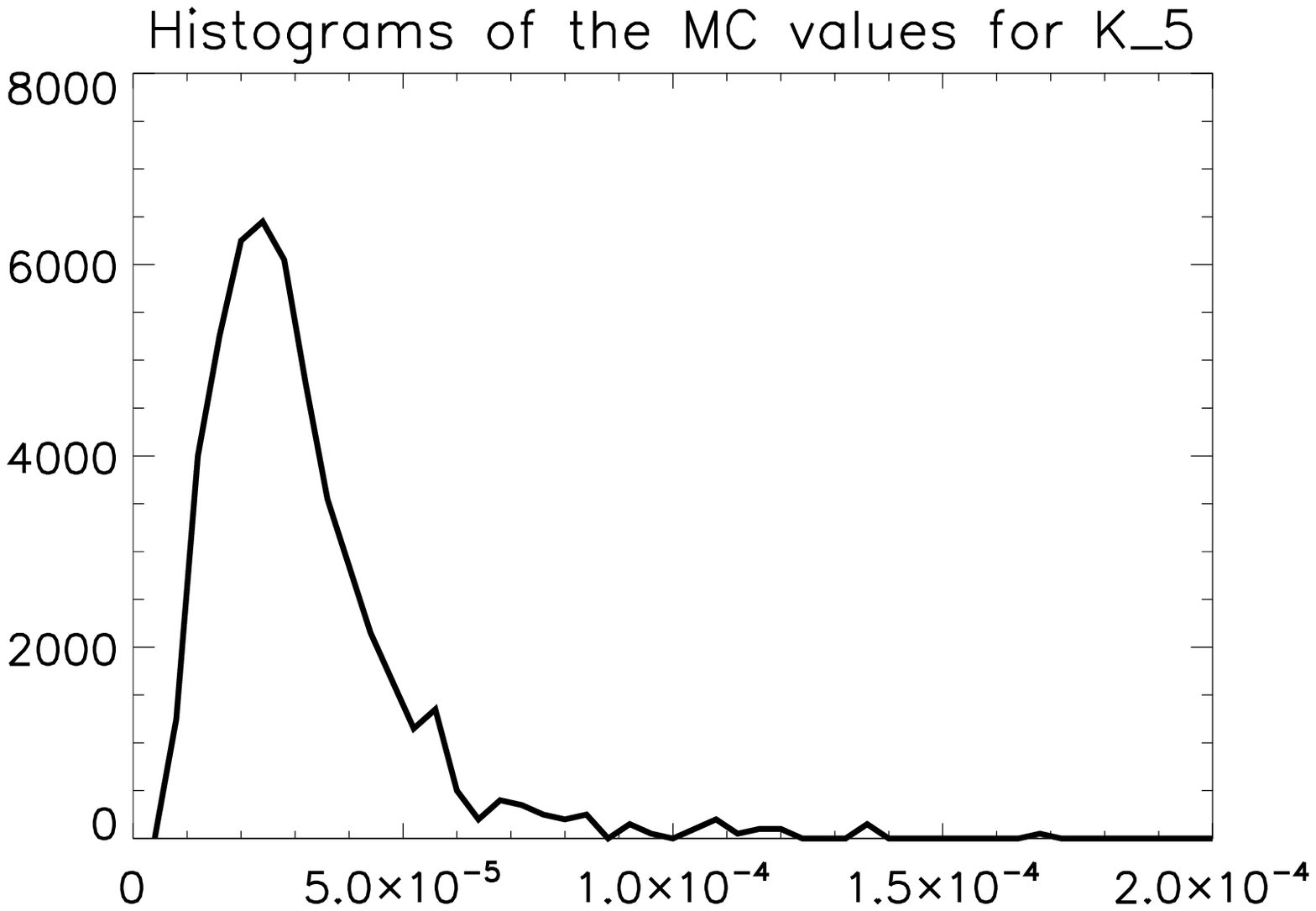}
\caption{\label{Fig4} Histograms show the 
distribution of the values for the multipoles $S_{5}$ and $K_{5}$ calculated
from the MC (\emph{scramb\/led}) ILC maps. This figure illustrates the
non-Gaussianity of a typical distribution for a given fixed $\ell$.}
\end{center}
\end{figure*}

To provide some additional qualitative information about the anisotropic
distribution of our non-Gaussianity indicators, we depict in Fig.~\ref{Fig3}
the dipole and the quadrupole as well as the full $S$ and $K$ maps with
these two components and the monopole removed. The left panels display
these components for the skewness indicator $S(\theta,\phi)$, while the
right panels show the same components for the kurtosis indicator
$K(\theta,\phi)$.
These multipoles maps were calculated from the WMAP five-year ILC masked data, but
corresponding maps for three-year WMAP data are largely similar to the
depicted maps, and were not included to avoid repetition.

It is known that the frequency K--band, Ka--band, Q--band, V--band, and W--band
foreground uncleaned maps, have different contaminants.
These features may appear in these maps in the form of non-Gaussianity even
when the \emph{KQ75} mask is utilized. 
Thus, in order to suitably test the  $S$ and $K$ indicators for non-Gaussianity
one should calculate their maps and angular power spectra not only for the
foreground reduced ILC masked and unmasked maps, but also for these single
frequency maps with and without \emph{KQ75} mask.
This also allows a comparative analysis of the outcomes.
To this end, we have also calculated the $S-$maps and $K-$maps for each
of these five frequency maps with and without \emph{KQ75} mask, for caps
with aperture $\gamma = 90^{\circ}$, centered at $N_\text{c}=12\,288$
points homogeneously generated on the sphere  by using  HEALPix.
However, to avoid repetition of figures which give only qualitative
information, in the following we shall concentrate on their angular power
spectra, which provide quantitative information.

\begin{figure*}[htb!]
\begin{center}
\includegraphics[width=8.8cm,height=5.6cm]{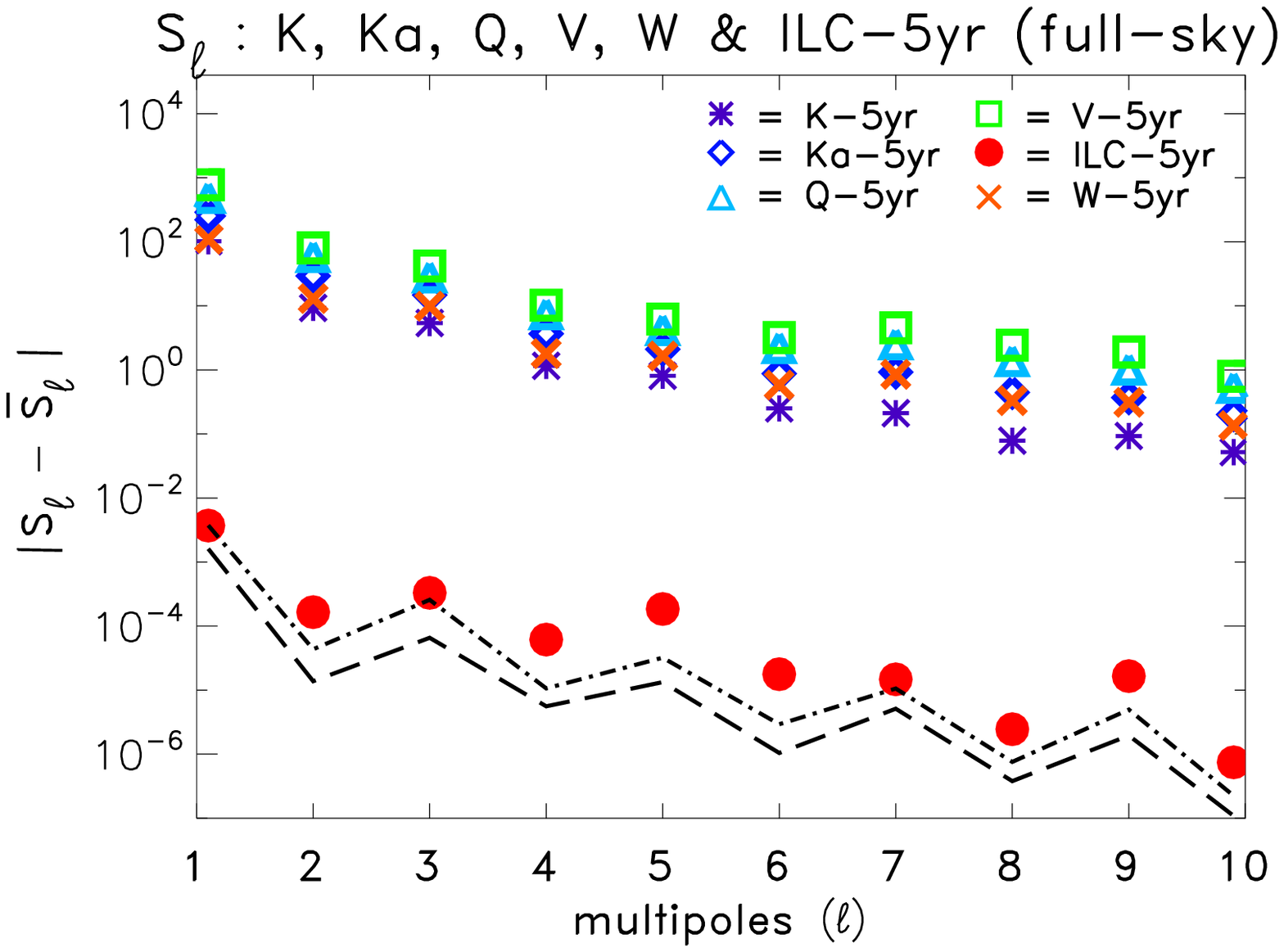}
\includegraphics[width=8.8cm,height=5.6cm]{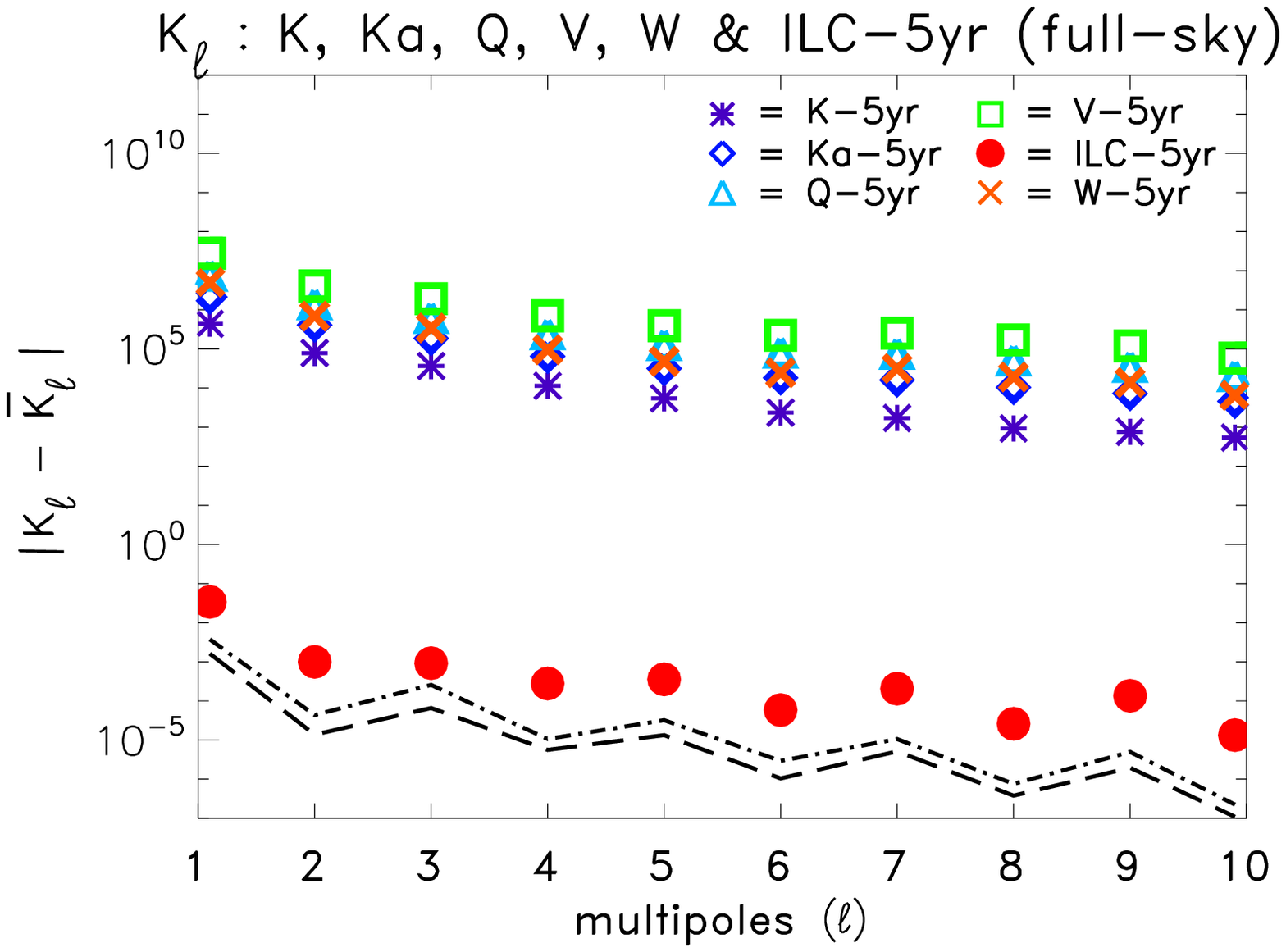}
\caption{\label{Fig5} Differential power spectrum of skewness
$|S_{\ell} - \overline{S}_{\ell}|$ and $|K_{\ell} - \overline{K}_{\ell}|$
(left) and kurtosis (right) indicators calculated from the and five-year
(right panels) WMAP CMB maps with no mask. The $68\%$ and $95\%$ confidence
levels are indicated, respectively,  by the dashed and dash-dotted lines.}
\end{center}
\end{figure*}

We calculated the angular power spectrum of the $S$ and $K$ maps generated from
the three and five-year data of ILC, and the foreground unreduced  K, Ka, Q, V and
W  maps with and without \emph{KQ75} mask.
These power spectra allow to estimate the statistical significance
of $S_{\ell}$ and $K_{\ell}$ by comparing them with the mean
angular power spectrum of the $S$ and $K$ maps obtained from
$1\,000$ Monte-Carlo-generated (MC) statistically Gaussian CMB maps.%
\footnote{We note that each MC map is a stochastic realization of the WMAP
best-fitting angular power spectrum of the $\Lambda$CDM model, obtained
by randomizing the multipole temperature components $a_{\ell m}$ within
the cosmic variance limits~\cite{wmap5,wmap3-Spergel-et-al}.}
To make easier this comparison, instead of using the angular power spectra
$S_{\ell}$ and $K_{\ell}$ themselves, we  employed
the \emph{differential} power spectra 
$|S_{\ell} - \overline{S}_{\ell}|$ and
$|K_{\ell} - \overline{K}_{\ell}|$.
Throughout the paper the mean quantities are denoted by overline.

To describe with some details our calculations we focus on the
skewness indicator $S$, since similar procedure holds 
for the kurtosis indicator $K$.
Starting from a given CMB seed map (ILC or any frequency band map) we 
generated $1\,000$ MC Gaussian (\emph{scramb\/led}) CMB maps, 
which are then used to generate  $1\,000$ 
skewness $S-$maps, from which we calculate $1\,000$ power 
spectra: $S^{\,\mathbf{i} }_{\,\ell}$ with enumeration index $\,\mathbf{i}
= 1,\,\,\cdots,1\,000\,$. In this way, for each fixed multipole component 
$S^{\,\mathbf{i} }_{\ell=\text{fixed}}$ we have $1\,000\,$ values 
from which we calculate the mean value
$\overline{S}_{\ell} = (1/1000) \sum_{\mathbf{i} = 1}^{1000} S_{\,\ell}^{\,\mathbf{i}}\,$. 
{}From this MC process we have at the end ten mean values 
$\overline{S}_{\ell}$ each of which are then used to compare with the
corresponding power spectrum component $S_{\ell}$ of the skewness map
obtained from an \emph{unscramb\/led} seed temperature WMAP map, in order to 
evaluate the statistical significance of each multipole component. 
Thus, for example, to study the statistical significance of the dipole 
moment of the skewness-ILC-map $S_{1}^{\,\text{\sc ILC}}$ we calculate
$|S_{1}^{\,\text{ILC}} - \overline{S}_{1}|$, where the mean dipole value
$\overline{S}_{1}$ is calculated 
from the $\,\mathbf{i}= 1,\,\,\cdots,1\,000\,$ power spectra of the Gaussian (\emph{scramb\/led}) maps.

The panels of Fig.~\ref{Fig4} display the histograms which show the 
distribution of the values for $S_5$ and $K_5$ calculated
from the \emph{scramb\/led} ILC maps. This figure makes clear that
a typical distribution of MC values for a given fixed $\ell$ is highly 
non-Gaussian. 

Figure~\ref{Fig5} shows the differential power spectra calculated from 
\emph{full-sky} CMB five-year maps, i.e., it displays the absolute value 
of the deviations from the mean angular power spectrum of the skewness $S_{\ell}$
(left panel) and kurtosis $K_{\ell}$ (right panel) indicators for
$\,\ell=1,\,\,\cdots,10\,$, which is a range of multipoles values useful to
investigate the large-scale angular characteristics of the $S$ and $K$ maps.
This figure makes apparent the strong deviation from Gaussianity of the
unmasked frequency maps. This is expected from the very outset since these
full-sky band maps are highly contaminated.  Figure~\ref{Fig5} also shows a
significant deviation from Gaussianity in five-year ILC unmasked
data, i.e. the deviations $|S_{\ell} - \overline{S}_{\ell}|$ and
$|K_{\ell} - \overline{K}_{\ell}|$ for the five-year ILC unmasked data 
are not within $95\%$  of the MC value. Actually the values of $S_{\ell}$ 
and $K_{\ell}$ obtained from the data are far beyond ($\gg 95\% $ off) the mean MC values 
(see Fig.~\ref{Fig5}). 
These results quantify and make clear the suitability of the WMAP team
recommendation to employ the new mask \emph{KQ75} for tests of
Gaussianity.
These spectra maps were calculated from the WMAP five-year data, but
the corresponding spectra for three-year data are very similar to the
depicted spectra. 
We stress that for the frequency maps we have used the foreground
uncleaned maps as the temperature fluctuations seed maps.

\begin{figure*}[tbh!]
\begin{center}
\includegraphics[width=8.8cm,height=5.6cm]{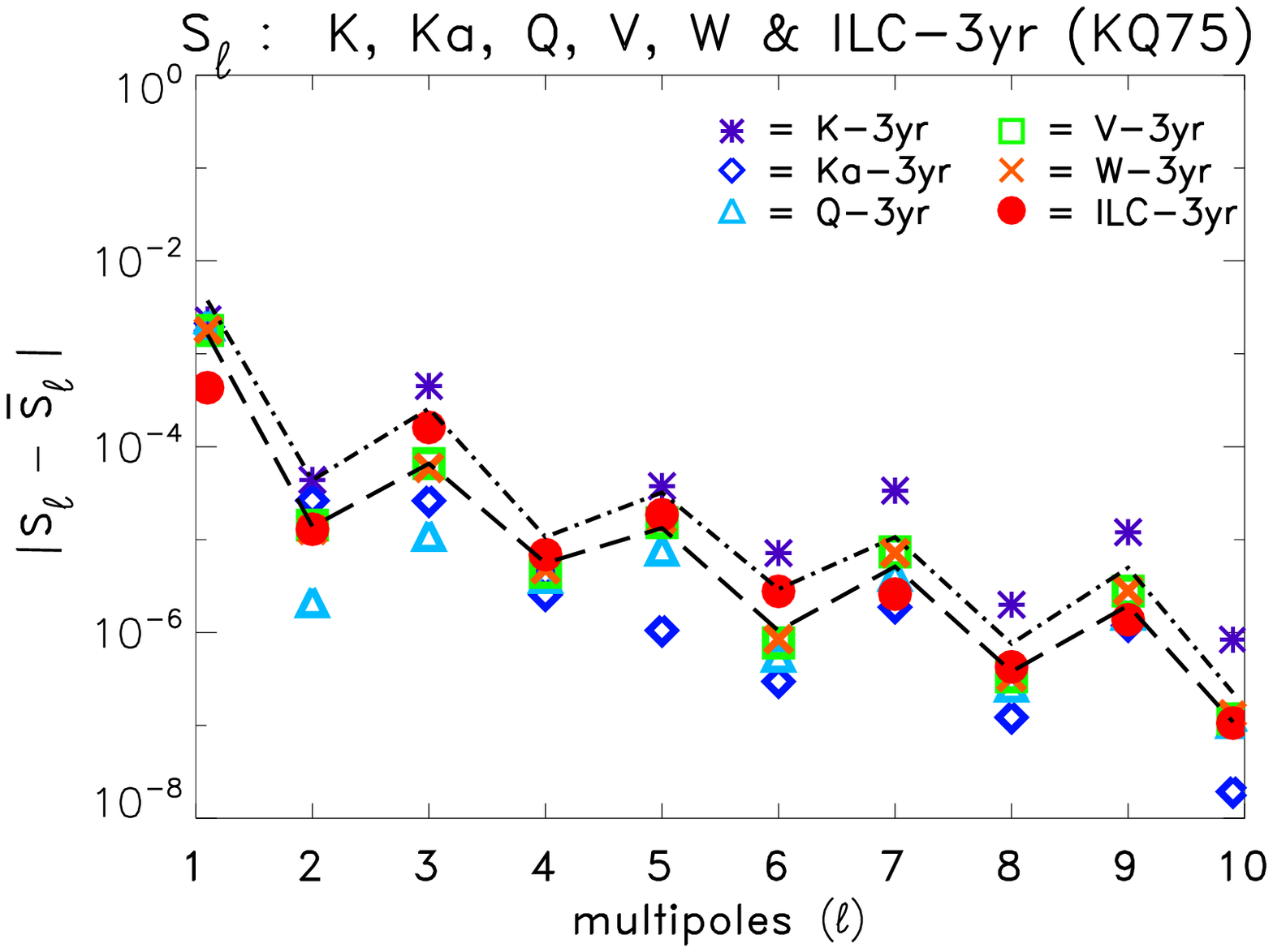}
\includegraphics[width=8.8cm,height=5.6cm]{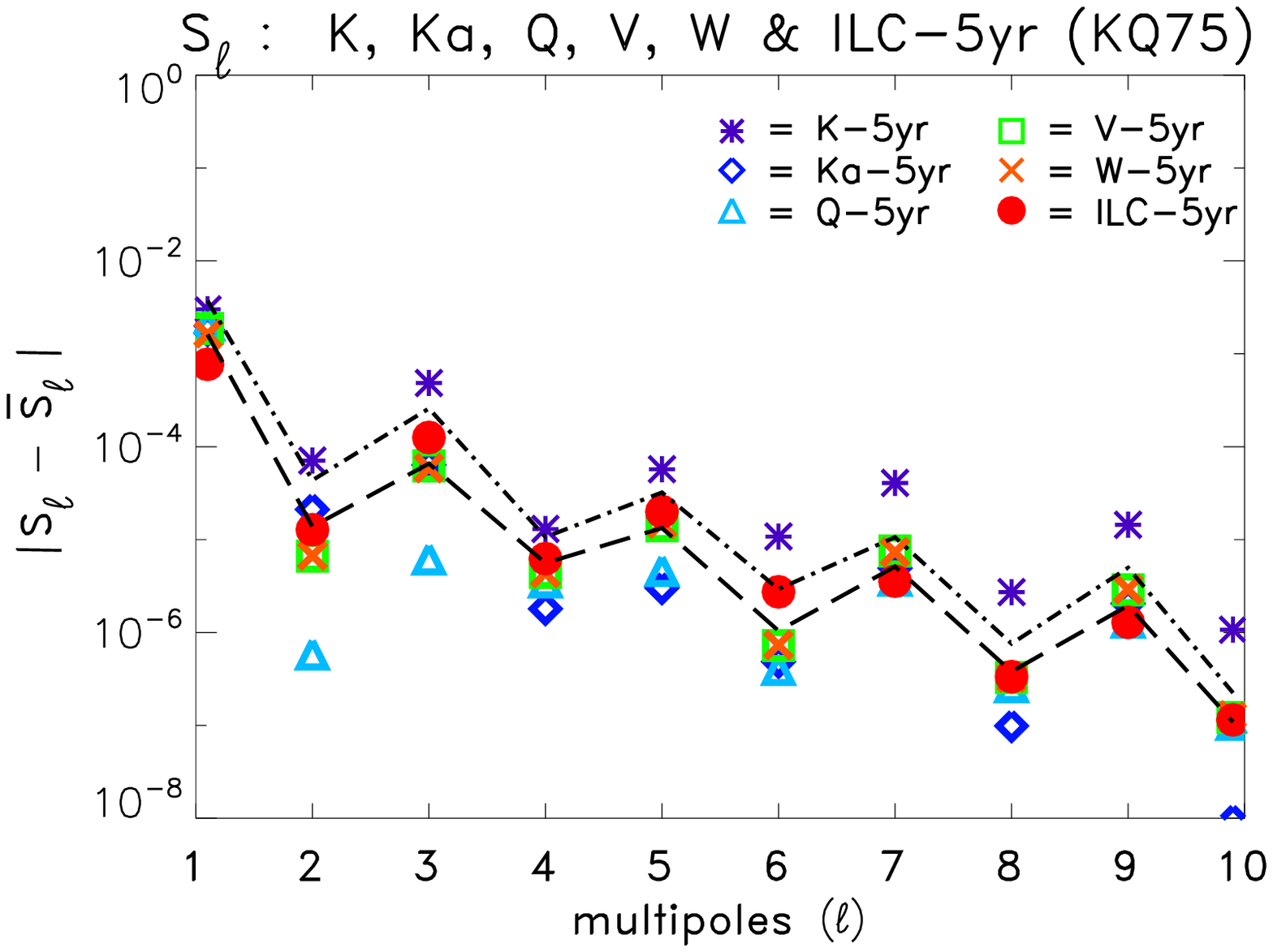}
\vskip 5mm
\includegraphics[width=8.8cm,height=5.6cm]{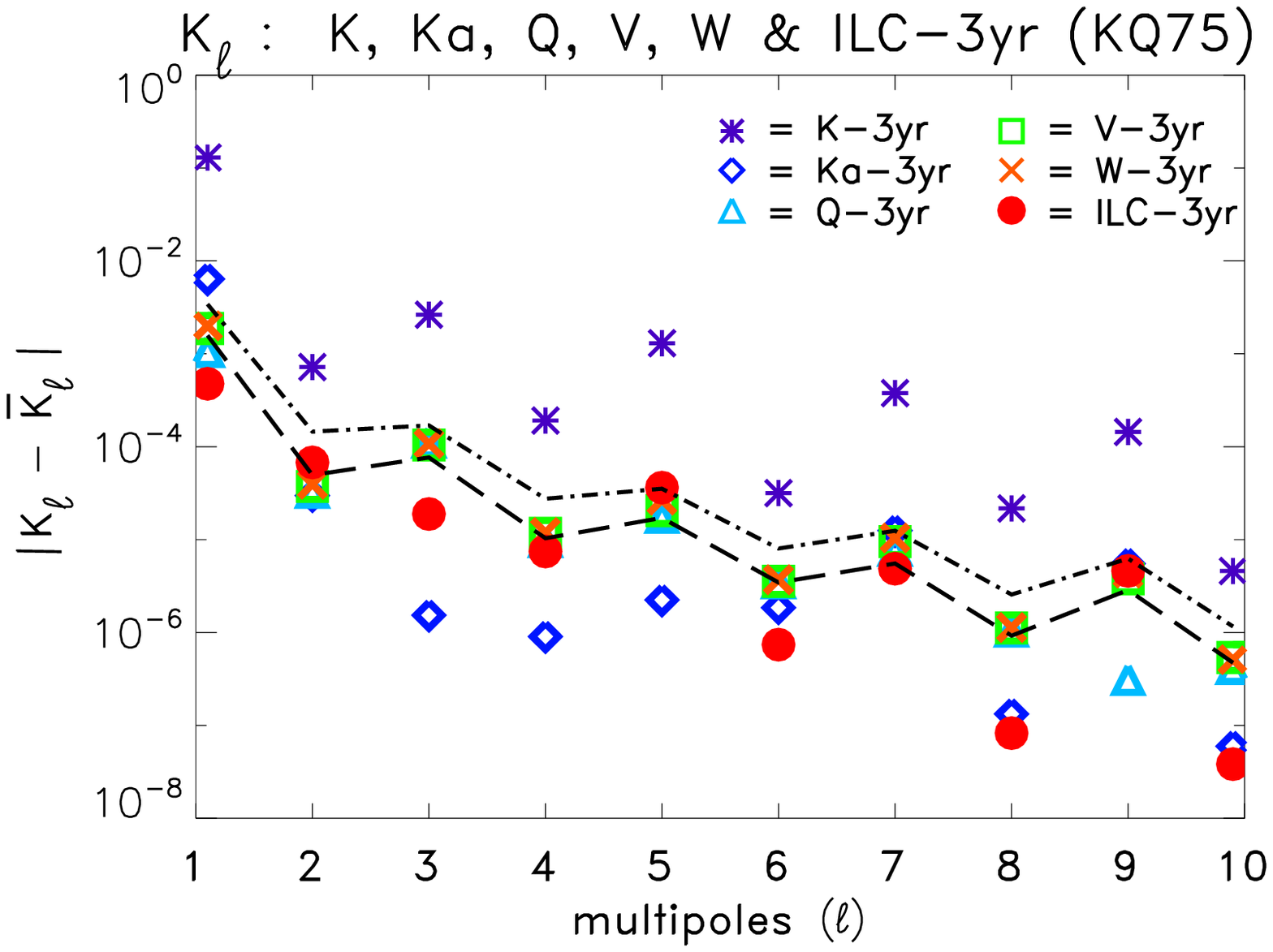}
\includegraphics[width=8.8cm,height=5.6cm]{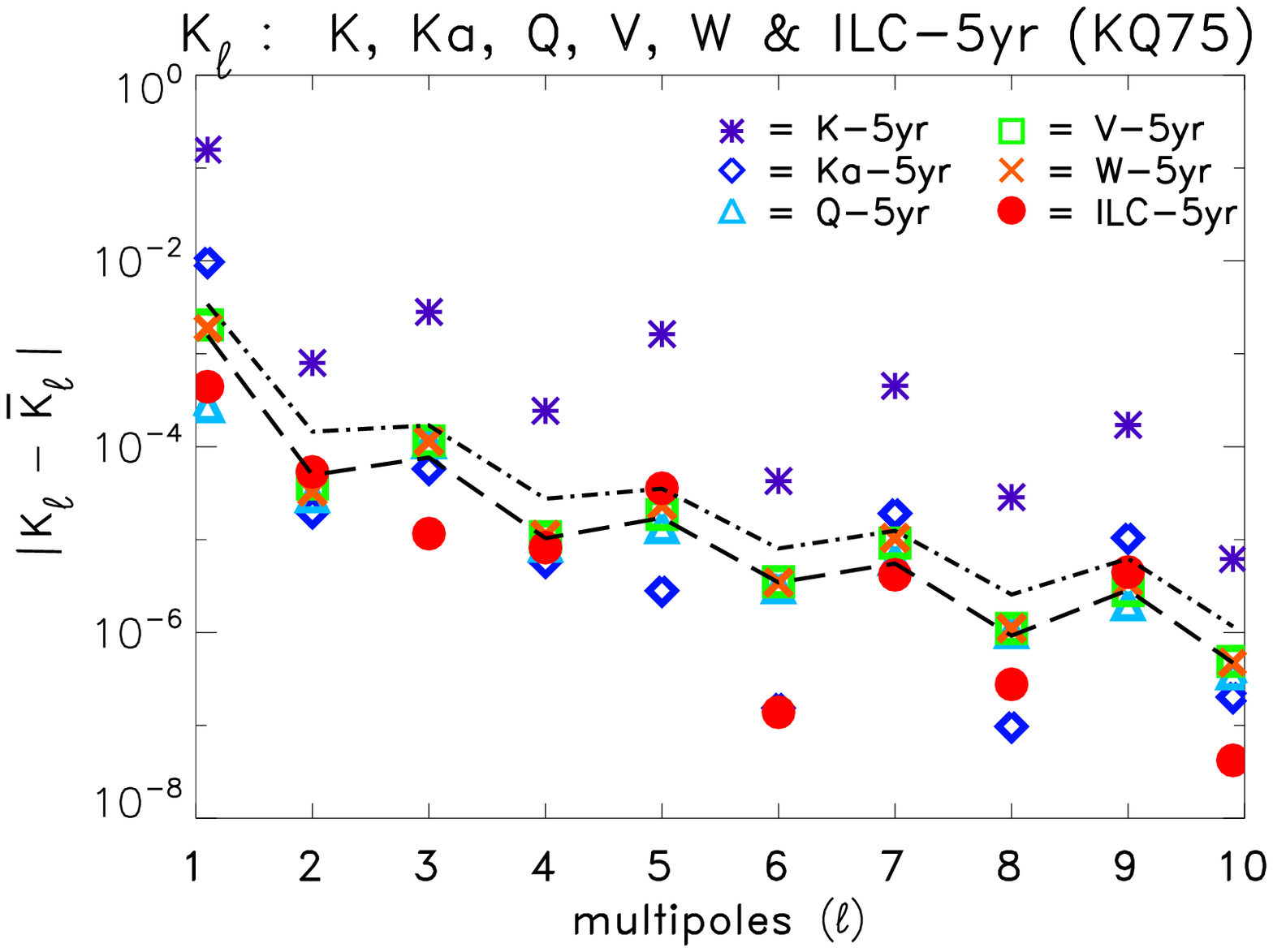}
\caption{\label{Fig6} Differential power spectrum of skewness
(first row) and kurtosis (second row) indicators calculated from the three
(left panels) and five-year (right panels) WMAP CMB maps with mask \emph{KQ75}.
The $68\%$ and $95\%$ confidence levels are indicated,  respectively,  by the 
dashed and dot-dashed lines. See the text for more details.}
\end{center}
\end{figure*}

Figure~\ref{Fig6} shows similar differential power spectra but now
calculated from the CMB three and five-year maps with \emph{KQ75} mask.
Tables~\ref{Skew-deviation} and~\ref{Kurt-deviation} complement Fig.~\ref{Fig6}
by collecting together the percentage of the deviations 
$|S^{\,\mathbf{i} }_{\,\ell} - \overline{S}_{\ell}|$ (calculated from
$1\,000$ \emph{scrambled} MC simulated  maps) which are smaller than
$|S_{\ell} - \overline{S}_{\ell}|$ obtained from the data, i.e. from the
five-year ILC, K, Ka, Q, V, and W masked maps. 
Thus, for example, according to Table~\ref{Skew-deviation} for the K band 
$99.6\%$ of multipoles $S^{\,\mathbf{i} }_{3}$ obtained from the  MC 
maps are closer to the mean $\overline{S}_{3}$ 
than the value $S_{3}$ calculated from the K map (three and five-year data). 
This indicates how unlikely (only $0.4\%$) is the occurrence of
the value obtained from the data for the multipole $S_{3}$ in the set of MC 
simulated maps for this band, giving therefore a clear indication of 
deviation from Gaussianity for the K masked maps to the extent that 
this  is not within $\sim 95\%$ of MC values. 

\begin{table*}[hbt]
\begin{center}
\begin{tabular}{ccccccc} 
\hline \hline 
$\ell$ &\ \ \ K \ &\ \ \  Ka \  &\ \ \ Q \ \ &\ \ \ V \ &\ \ \ W \ & \ \ \ ILC   \\
\hline  
$1$ & \ \ \ $90.6$ --- $93.2$ & \ \ \ $83.7$ --- $76.4$ & \ \ \ $88.5$ --- $88.7$ & \ \ \ $74.5$ --- $80.6$
& \ \ \ $76.0$ --- $66.4$ & \ \ \ $14.8$ --- $26.3$ \\
$2$ & \ \ \ $95.9$ --- $97.7$ & \ \ \ $92.3$ --- $90.1$ & \ \ \ $6.5$  --- $1.3$  & \ \ \ $70.5$ --- $22.8$
& \ \ \ $57.2$ --- $23.2$ & \ \ \ $59.0$ --- $57.9$ \\
$3$ & \ \ \ $99.6$ --- $99.6$ & \ \ \ $24.1$ --- $63.6$ & \ \ \ $10.0$ --- $5.7$  & \ \ \ $66.1$ --- $61.5$
& \ \ \ $57.7$ --- $57.5$ & \ \ \ $96.3$ --- $94.3$ \\
$4$ & \ \ \ $71.0$ --- $94.4$ & \ \ \ $23.7$ --- $16.0$ & \ \ \ $39.0$ --- $34.3$ & \ \ \ $47.0$ --- $46.6$
& \ \ \ $52.3$ --- $45.6$ & \ \ \ $90.0$ --- $83.7$ \\
$5$ & \ \ \ $96.4$ --- $98.4$ & \ \ \ $6.4$  --- $16.1$ & \ \ \ $38.8$ --- $21.9$ & \ \ \ $74.6$ --- $70.1$
& \ \ \ $80.8$ --- $72.4$ & \ \ \ $86.0$ --- $88.7$ \\
$6$ & \ \ \ $99.5$ --- $99.7$ & \ \ \ $16.7$ --- $30.1$ & \ \ \ $34.2$ --- $23.6$ & \ \ \ $45.8$ --- $42.5$
& \ \ \ $51.5$ --- $43.3$ & \ \ \ $96.2$ --- $96.0$ \\
$7$ & \ \ \ $99.7$ --- $99.7$ & \ \ \ $24.4$ --- $64.1$ & \ \ \ $53.0$ --- $46.7$ & \ \ \ $88.7$ --- $89.3$
& \ \ \ $87.5$ --- $88.6$ & \ \ \ $33.4$ --- $45.9$ \\
$8$ & \ \ \ $99.5$ --- $99.7$ & \ \ \ $19.0$ --- $16.0$ & \ \ \ $43.8$ --- $43.4$ & \ \ \ $59.4$ --- $57.7$
& \ \ \ $58.8$ --- $57.8$ & \ \ \ $83.7$ --- $59.4$ \\
$9$ & \ \ \ $99.6$ --- $99.7$ & \ \ \ $41.6$ --- $70.7$ & \ \ \ $53.4$ --- $45.4$ & \ \ \ $87.1$ --- $87.9$
& \ \ \ $87.2$ --- $88.2$ & \ \ \ $47.0$ --- $45.0$ \\
$10$& \ \ \ $99.8$ --- $99.8$ & \ \ \ $10.1$ --- $4.1$  & \ \ \ $67.5$ --- $63.4$ & \ \ \ $75.4$ --- $77.6$
& \ \ \ $82.9$ --- $81.9$ & \ \ \ $66.0$ --- $72.3$ \\
\hline
\end{tabular}
\end{center}
\caption{Percentage of the deviations $|S^{\,\mathbf{i} }_{\,\ell} - 
\overline{S}_{\ell}|$ (for $\,\ell=1,\,\,\cdots,10\,$ and 
calculated from $1\,000$ \emph{scrambled} MC simulated  maps), 
which are smaller than
$|S_{\ell} - \overline{S}_{\ell}|$ obtained from the data:  
$S-$maps generated from the three (left entries of each column) 
and five-year (right accesses for each column) CMB maps in the frequencies 
K ($22.8$ GHz), Ka ($33.0$ GHz), Q ($40.7$ GHz), V ($60.8$ GHz), and 
W ($93.5$ GHz), along with the ILC three and five-year maps.}
\label{Skew-deviation}
\end{table*}

\begin{table*}[!bht]
\begin{center}
\begin{tabular}{ccccccc} 
\hline \hline 
$\ell$ &\ \ \ K \ &\ \ \  Ka \  &\ \ \ Q \ \ &\ \ \ V \ &\ \ \ W \ & \ \ \ ILC
\\
\hline  
$1$ & \ \ \ $99.9$ --- $99.9$ & \ \ \ $98.7$ --- $99.3$ & \ \ \ $44.5$ --- $11.7$ & \ \ \ $82.0$ --- $87.8$
& \ \ \ $86.4$ --- $82.6$ & \ \ \ $19.4$ --- $18.9$ \\
$2$ & \ \ \ $99.9$ --- $99.9$ & \ \ \ $27.1$ --- $14.1$ & \ \ \ $29.8$ --- $25.2$ & \ \ \ $38.4$ --- $42.1$
& \ \ \ $41.6$ --- $30.3$ & \ \ \ $74.9$ --- $76.4$ \\
$3$ & \ \ \ $99.9$ --- $99.9$ & \ \ \ $1.7$  --- $51.5$ & \ \ \ $90.1$ --- $89.5$ & \ \ \ $86.9$ --- $90.9$
& \ \ \ $88.7$ --- $90.8$ & \ \ \ $16.0$ --- $10.9$ \\
$4$ & \ \ \ $99.8$ --- $99.8$ & \ \ \ $3.0$  --- $24.8$ & \ \ \ $55.4$ --- $45.6$ & \ \ \ $84.8$ --- $72.3$
& \ \ \ $89.1$ --- $79.6$ & \ \ \ $37.2$ --- $42.9$ \\
$5$ & \ \ \ $99.9$ --- $99.9$ & \ \ \ $7.9$  --- $10.5$ & \ \ \ $69.4$ --- $54.3$ & \ \ \ $80.1$ --- $74.1$
& \ \ \ $93.2$ --- $87.6$ & \ \ \ $94.7$ --- $95.3$ \\
$6$ & \ \ \ $99.8$ --- $99.8$ & \ \ \ $21.0$ --- $2.2$  & \ \ \ $66.1$ --- $49.2$ & \ \ \ $73.0$ --- $71.7$
& \ \ \ $80.1$ --- $67.4$ & \ \ \ $8.3$  --- $1.9$  \\
$7$ & \ \ \ $99.9$ --- $99.9$ & \ \ \ $95.1$ --- $97.9$ & \ \ \ $86.5$ --- $74.9$ & \ \ \ $92.5$ --- $91.8$
& \ \ \ $93.7$ --- $93.7$ & \ \ \ $60.3$ --- $53.1$ \\
$8$ & \ \ \ $99.8$ --- $99.8$ & \ \ \ $7.2$  --- $4.8$  & \ \ \ $82.8$ --- $76.9$ & \ \ \ $91.8$ --- $90.0$
& \ \ \ $93.2$ --- $90.7$ & \ \ \ $4.3$  --- $13.4$ \\
$9$ & \ \ \ $99.9$ --- $99.9$ & \ \ \ $94.8$ --- $97.9$ & \ \ \ $5.9$  --- $47.4$ & \ \ \ $85.6$ --- $68.8$
& \ \ \ $92.3$ --- $81.0$ & \ \ \ $93.5$ --- $95.0$ \\
$10$& \ \ \ $99.8$ --- $99.8$ & \ \ \ $3.2$  --- $17.1$ & \ \ \ $48.7$ --- $37.1$ & \ \ \ $89.2$ --- $73.3$
& \ \ \ $80.4$ --- $67.1$ & \ \ \ $2.2$  --- $2.3$  \\
\hline
\end{tabular}
\end{center}
\caption{Percentage of the deviations 
$|K^{\,\mathbf{i} }_{\,\ell} - \overline{K}_{\ell}|$ (calculated from
$1\,000$ \emph{scrambled} MC random  maps for $\,\ell=1,\,\,\cdots,10\,$), 
which are smaller than
$|K_{\ell} - \overline{K}_{\ell}|$ obtained from the data:  
$K-$maps generated from the three (left entries of each column) and five-year 
(right entries for each column) CMB seed maps in the channels K ($22.8$ GHz),  
Ka ($33.0$ GHz), Q ($40.7$ GHz), V ($60.8$ GHz), and W ($93.5$ GHz), 
along with the ILC three and five-year maps.}
\label{Kurt-deviation}
\end{table*}

\begin{figure*}[bht!]
\begin{center}
\includegraphics[width=5.2cm,height=8.2cm,angle=90]{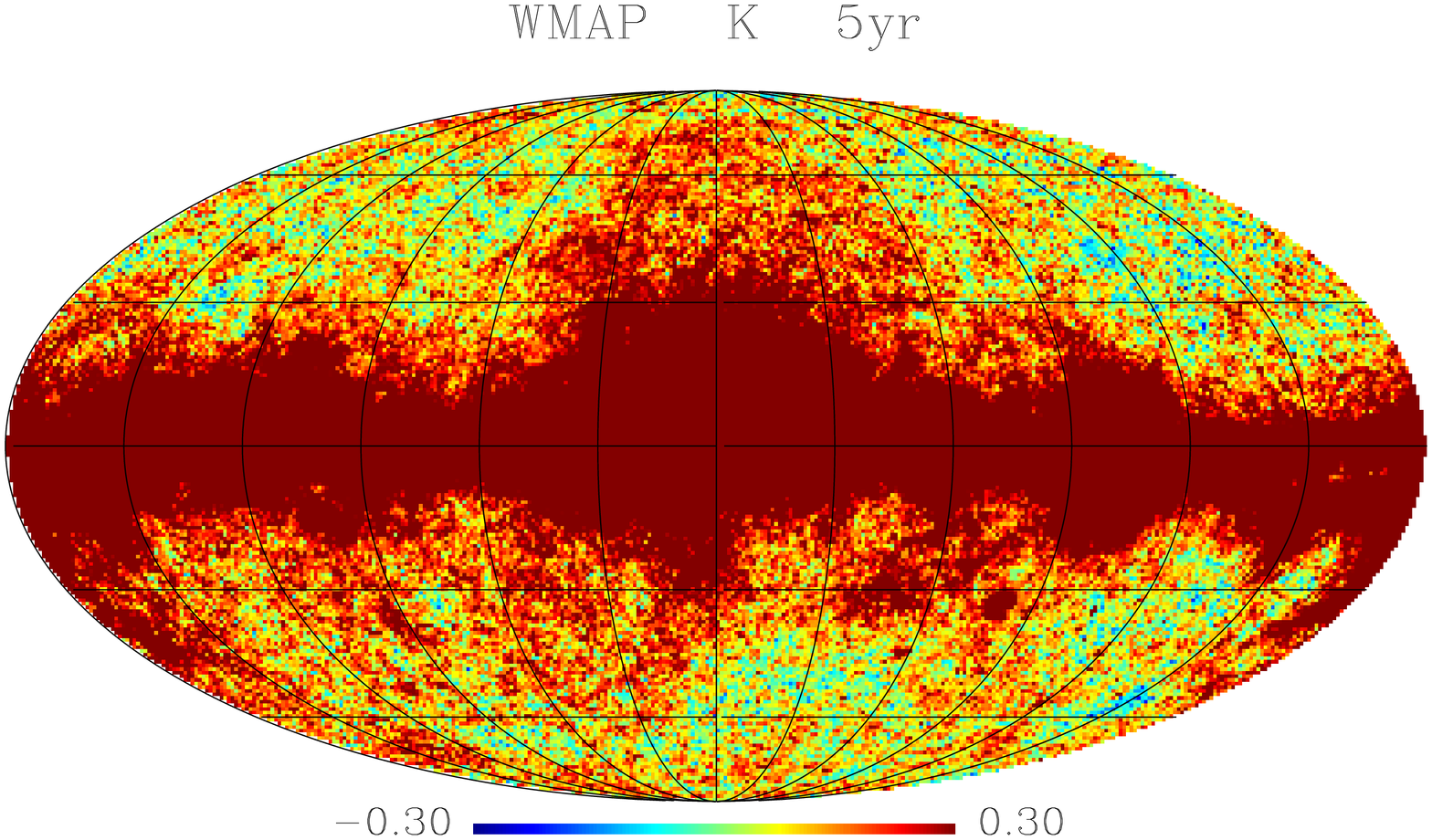} 
\includegraphics[width=5.2cm,height=8.2cm,angle=90]{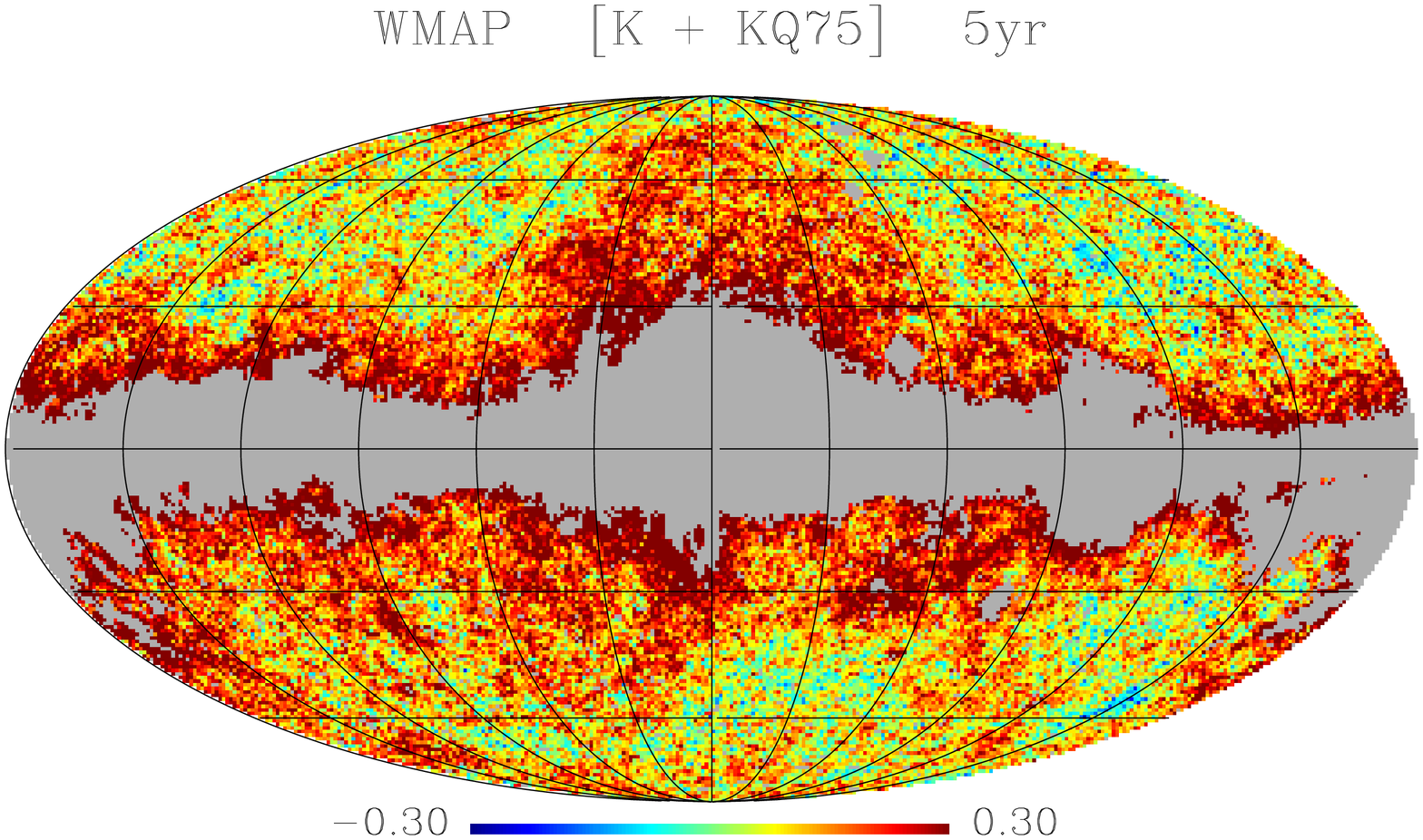} 
\includegraphics[width=5.2cm,height=8.2cm,angle=90]{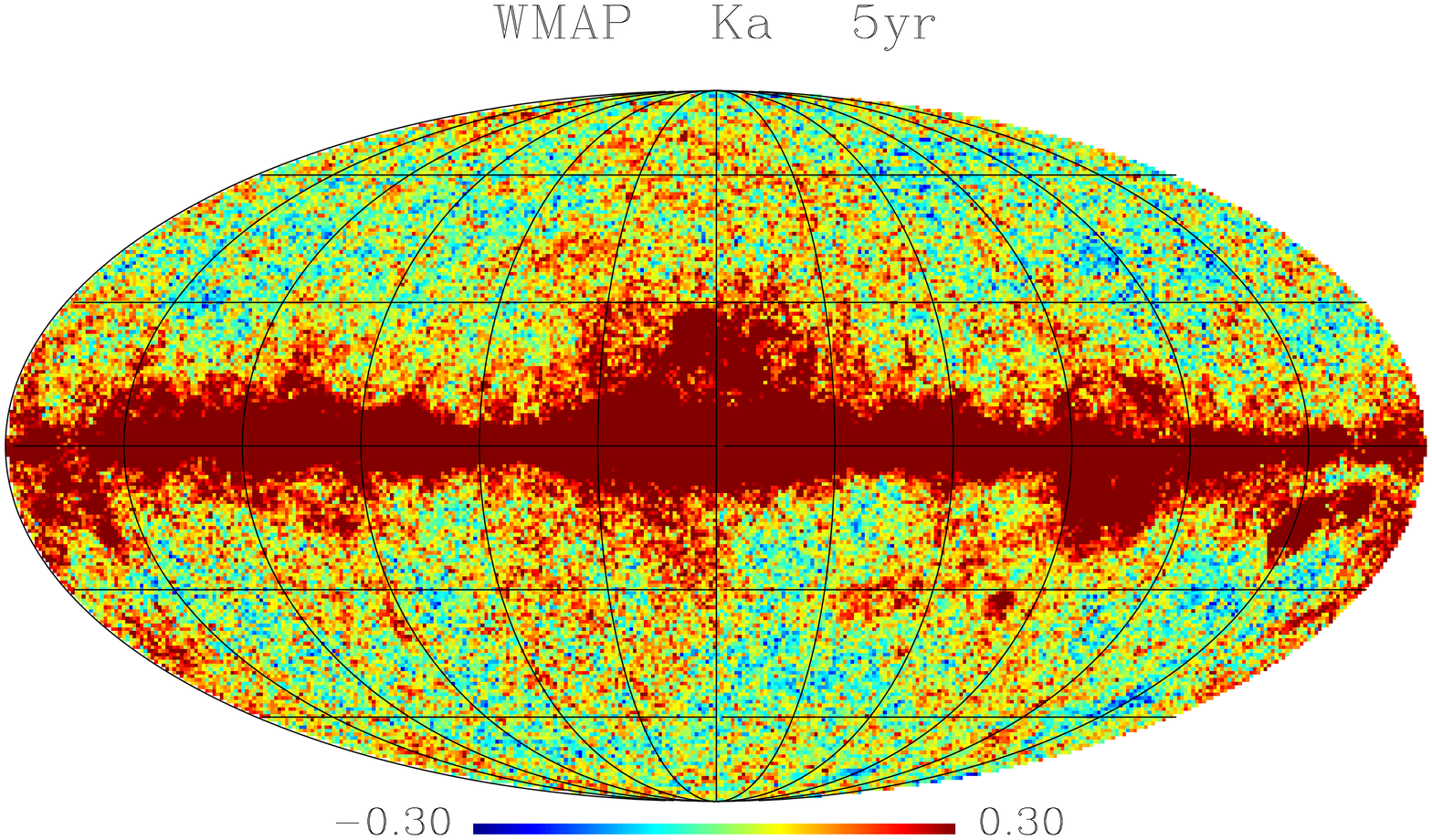} 
\includegraphics[width=5.2cm,height=8.2cm,angle=90]{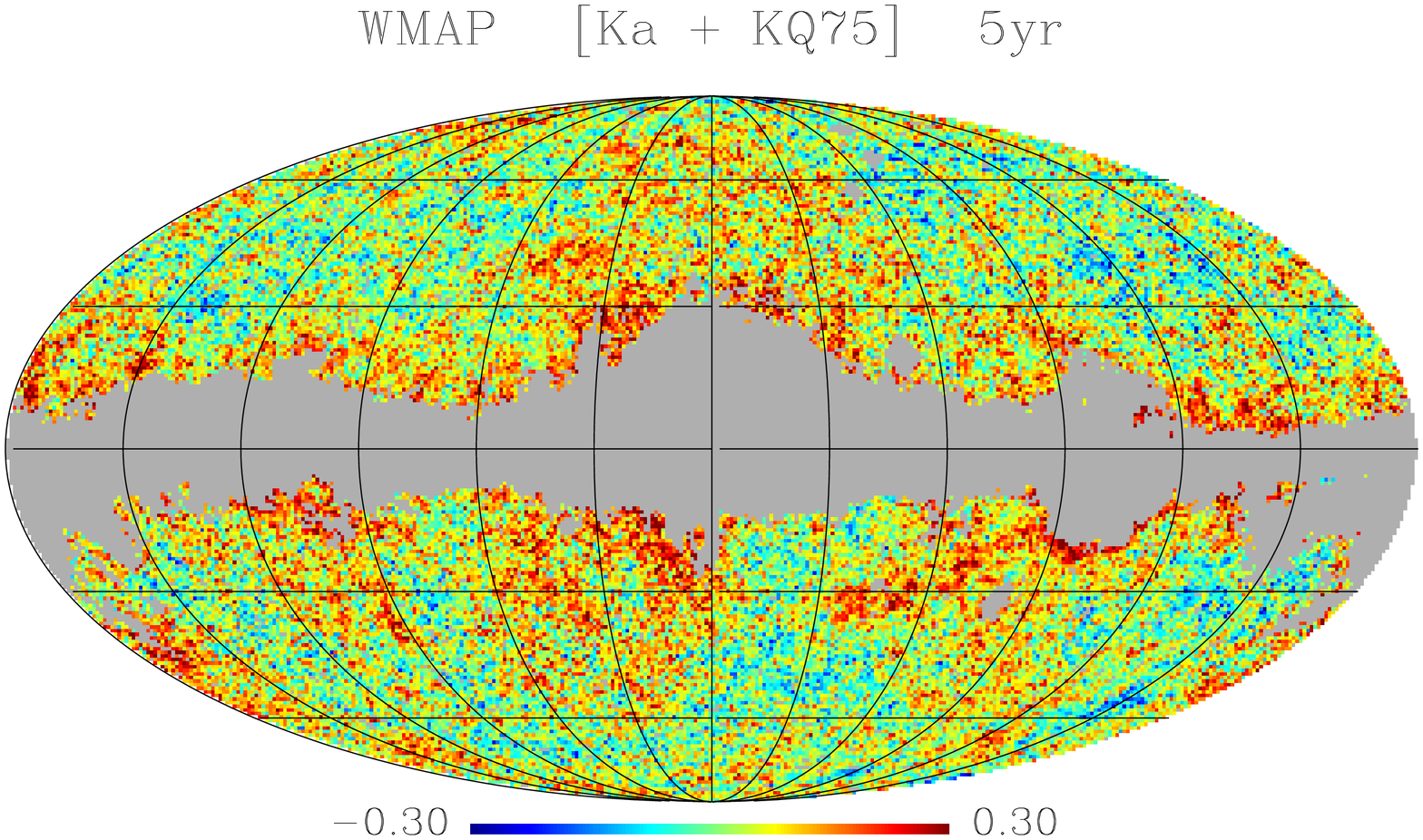} 
\caption{\label{With_without} CMB temperature fluctuations five-year maps
in the frequencies K ($22.8$ GHz) and Ka ($33.0$ GHz) without (left panel)
and with the mask \emph{KQ75}. These figure illustrated that
even these masked maps are still foreground contaminated at some level.
These signs of deviation from Gaussianity in the \emph{KQ75} masked maps
are captured by our indicators $S$ (K \emph{KQ75} masked map) and $K$
(K and Ka \emph{KQ75} masked maps).}
\end{center}
\end{figure*}

Regarding the angular power spectrum of the skewness and kurtosis maps calculated
from the ILC  maps with \emph{KQ75} mask, it is clear from the
Fig.~\ref{Fig6} and Tables~\ref{Skew-deviation} and~\ref{Kurt-deviation}
that the low $\ell$ multipoles ($\,\ell=1,\,\,\cdots,10\,$) values 
--- indicative of large-angle deviation from Gaussianity ---  are not
statistically significant, i.e. they are within the $\sim 95\%$  of the
MC values. In other words, the occurrence of these ILC multipole 
values is very likely ($\geq 5\%$) in the set of MC values.%
\footnote{We note that the values of $S_6$ (five-year) and 
$K_5$ (five-year) which are slightly smaller than $5\%$.}
It is interesting to note that a comparison between Fig.~\ref{Fig5} 
and Fig.~\ref{Fig6} makes apparent the important role of the mask in 
the test for Gaussianity carried out.
Furthermore, Fig.~\ref{Fig6} and Tables~\ref{Skew-deviation}
and~\ref{Kurt-deviation} also show that this analysis
is robust with respect to the three and five-year ILC maps.

As regards the angular power spectra of $S$ and $K$ calculated from the K--band
map with \emph{KQ75} mask, it is clear from Fig.~\ref{Fig6} and
Tables~\ref{Skew-deviation} and~\ref{Kurt-deviation}  that both the skewness
and kurtosis spectra reveal deviations from the mean power spectrum values,
greater than  $95\%$, for $\ell = 2, 3, 5, 6, 7, 8, 10$.
This indication of deviation from Gaussianity  comes chiefly from the 
fact that even with \emph{KQ75} masked K--band map is still foreground contaminated 
as illustrated in first row [panels (a) and (b)] of Fig.~\ref{With_without}.

Figure~\ref{With_without} [second row, panels (c) and (d)] suggests a residual
foreground contamination in the Ka--band map with \emph{KQ75} mask. This
remnant contamination from galactic foreground emission is detected by the
kurtosis indicator $K$, whose power spectrum shows an excess of power relative to
the mean for $\ell = 1, 7, 9$ components [see the second row of panels in  
Fig.~\ref{Fig6} and Table~\ref{Kurt-deviation}].

Regarding the angular power spectra of the $S$ and $K$ indicators for the
remaining three and five-year frequency maps (Q, V, W)  with \emph{KQ75} mask
the panels of Fig.~\ref{Fig6} along with Tables~\ref{Skew-deviation}
and~\ref{Kurt-deviation} show that the occurrence of the power spectrum 
low $\ell$ values calculated from these frequency maps  are very likely 
($>5\%$) in the set of MC values, i.e. the Q, V and W multipoles values 
are well within $\simeq 95\%$ of MC values.
This makes clear that for these frequency maps the \emph{KQ75} mask 
reduces considerably the contamination to a level which make the
power spectra  of this masked map compatible with Gaussianity. 

These results of our statistical analyses indicate that
the current CMB temperature fluctuations of the ILC
foreground reduced three and five-year \emph{KQ75} masked maps
are consistent with Gaussianity in that the occurrence of these ILC multipole 
multipole values is very likely ($\geq 5\%$) in the set of MC values, i.e.
the ILC multipole values of the $S$ and $K$ are within $\simeq 95\%$  of 
the MC values of the \emph{scrambled} (Gaussian) maps. This result agrees with 
the WMAP team and other analyses made by using different statistical
tools~\cite{wmap1,wmap3,wmap5,Gauss-consist}.
On the other hand, our analyses detect a deviation from Gaussianity in
the three and five-year frequency \emph{KQ75} masked K and Ka maps
which is consistent with the fact that even these masked maps are
foreground contaminated at some level (see Fig.~\ref{With_without}).
As for the Q, V, and W  frequency unremoved foreground maps, our
analyses indicate that although considerably contaminated in their
full-sky form (as captured in Fig.~\ref{Fig5}), \emph{KQ75} mask
cuts the galactic regions so as to bring their $S$ and $K$ power
spectra to a level consistent with Gaussianity (the multipoles values 
are within $\sim 95\%$ of MC values).

Finally, to have an overall assessment of the power spectra of the $S$ and $K$ 
maps calculated from each CMB seed map we have performed a $\chi^2$ test to 
find out the goodness of fit for $S_{\ell}$ and $K_{\ell}$ multipole values 
as compared to the expected multipoles values from the Gaussian MC maps. 
In this way we can obtain one number that collectively quantifies the extent 
to which a given \emph{KQ75} masked map is consistent with Gaussianity. 
For the power spectra $S_\ell$ we found that the ratio $\chi^2/\,\text{dof}\,$ 
(dof stands for degree of freedom) from the K, Ka, Q, V, W and ILC maps are given,
respectively, by $21.5$, $4.9$, $6.0$, $5.2$, $3.9$ and $1.2$, while for
the kurtosis power spectra $K_{\ell}$ of these maps the values of $\chi^2/\text{dof}$
are, respectively,  $35\,652$, $135$, $0.5$, $6.4$, $5.6$ and $0.4$. 
Clearly the greater are these values the smaller are the $\chi^2$ probability, i.e., 
the probability that the multipoles values $S_{\ell}$ and $K_{\ell}$ (from each CMB 
maps) and the expected MC multipole values agree.  A cut-off 
value of $\chi^2/\text{dof} \simeq 7$ to reject the the hypothesis 
that the multipole values $S_{\ell}$ and $K_{\ell}$  
are a good approximation to the expected 
multipole values of MC Gaussian realizations, is therefore enough
to ensure that  Q, V, W and ILC \emph{KQ75} masked are consistent
with Gaussianity, while the K and Ka   (\emph{KQ75} masked) maps
are not.

The calculations of our non-Gaussianity indicators require not only
the choice of a CMB map as input, but also the specification of
some quantities whose choice could in principle affect the outcome
of our results.
To test the robustness of our scheme, hence of our results, we
studied the effects of changing various parameters employed in
the calculation of our indicators.
We found that the $S$ and $K$ angular power spectra do not
change appreciably as we change the WMAP ILC three-year and
five-year maps; the resolution of CMB temperature maps used
($786\,432$ or $3\,145\,728$ pixels); and the number of point-centers
of the caps with values $768$, $3\,072$ and $12\,288$. The deviations
from the mean power spectrum for three-year and five-year frequency
bands maps are also robust (see Table~\ref{Skew-deviation} and
Table~\ref{Kurt-deviation}).

Concerning the robustness of the above analyses some additional words of
clarification are in order here. First, we note that the calculations of
the $S-$maps and $K-$maps by scanning the CMB maps sometimes include caps
whose center is within or close to the \emph{KQ75} masked region.
In these cases, the calculations of these indicators are made with smaller
number of pixels, which clearly introduce additional statistical noise
as compared to the cases whose caps centers are far away from the mask.
In order to minimize this effect  we have scanned the CMB sky with spherical
caps of aperture  $\gamma = 90^{\circ}$.%
\footnote{Note that even in the case of $90^{\circ}$ caps the sky regions
whose center are within or close to the galaxy have smaller sky fraction than
those centered far away from the mask. This could lead to bias in our
non-Gaussianity estimators. A possible way to circumvent this problem is
by dividing $S_j$ and $K_j$ by the corresponding standard deviation
$(\overline{S_j^2} - \overline{S_j}^2)^{1/2}$
or $(\overline{K_j^2} - \overline{K_j}^2)^{1/2}$,
obtained by using MC CMB seed maps. We have calculated the maps and power
spectra for these normalized indicators for $1\,000$ MC maps.
It turns out that for $\gamma = 90^{\circ}$ caps,
the power spectra of normalized and unnormalized indicators,
calculated for the  WMAP ILC five-year map, essentially coincide.
For caps of smaller aperture as, e.g., $\gamma = 60^{\circ}$,
although the normalization gives rise to smaller powers 
for $S_{\ell}$ and $K_{\ell}$, the loss of powers for each $\ell$ is
not enough to make the $\gamma = 60^{\circ}$ results close to the
clearly statistically less noisy $\gamma = 90^{\circ}$ power spectrum
estimates.}

Second, we have checked the sensitivity of our results regarding ILC maps
with respect to the angular resolution of CMB maps, the number $N_\text{c}$ of
point-centers of the caps used to scan the CMB sky, and the aperture $\gamma$
of the caps, by carrying out correlation analyses between the resultant
$S$ and $K$ maps calculated for CMB maps with $3\,145\,728$ and $786\,432$
pixels along with $N_\text{c} = 768$, $3\,072$ and $12\,288$.
We found that for $\gamma = 90^{\circ}$ the resulting $S$ and $K$ maps
are strongly correlated according to Pearson's correlation coefficient.
Thus, for example, for the $S-$maps calculated from the two CMB pixelizations
the Pearson's correlation coefficients are equal to $99.995\%$, $99.995\%$ and
$99.996\%$ for $N_\text{c} = 768$, $3\,072$ and $12\,288$, respectively, while for
the $K-$maps this coefficient is $99.996\%$ for the three values of $N_\text{c}$.
However, for caps with aperture $\gamma=70^{\circ}$, for example, the Pearson's
correlation coefficients for $S$ and $K$ maps (calculated for CMB maps with
these two pixelizations and the same values of $N_\text{c}$) fall below $50\%$, making
clear that caps of aperture  $\gamma = 90^{\circ}$ are the most suitable
spherical caps to deal with the statistical noise in our scheme.

\section{Final remarks and conclusions}
\label{Conclusions}

We have proposed two new indicators for measuring large-angle directional 
deviations from Gaussianity in the CMB data, and have used them to search 
for the large-angle (low $\ell$) deviation from Gaussianity in the three and
five-year foreground reduced ILC and the five single frequency (K, Ka, Q, V, W)
\emph{KQ75} masked maps. 
Our directional indicators enable us to construct skewness $S$ and kurtosis
$K$ maps (as, e.g., Fig.~\ref{Fig1} and Fig.~\ref{Fig2}), making it possible to
examine the presence and significance of possible large-angle non-Gaussianity
in the WMAP CMB temperature fluctuations maps with and without \emph{KQ75} mask. 

To obtain a more quantitative measure of non-Gaussianity we have studied
the low $\ell$ angular power spectrum of the $S$ and $K$ maps generated
from the (three and five-year) WMAP ILC and frequency (unremoved foreground)
maps with and without \emph{KQ75} mask. For the full-sky ILC maps we found
deviation from Gaussianity for the low $\ell$, while
for the ILC masked maps we found that the low $\ell$ ($\,\ell=1,\,\,\cdots,10\,$)
components are not significantly different from corresponding components
of the expected power spectrum calculated from $S$ and $K$  maps
obtained from $1\,000$ Monte Carlo CMB maps generated by considering
the Gaussian random hypothesis based on the concordance model~\cite{wmap5}.
Actually, we have found that the values of
the multipoles $S_{\ell}$ and $K_{\ell}$ (for $\,\ell=1,\,\,\cdots,10\,$)
are not statistically significant, i.e. they are within the $95\%$ values
of $S_{\ell}$'s and $K_{\ell}$'s of the MC randomly scrambled maps]
(Fig.~\ref{Fig6} and Tables~\ref{Skew-deviation} and~\ref{Kurt-deviation}).
As regards the frequency maps, we found clear indication of deviation from 
Gaussianity in the three and five-year frequency \emph{KQ75} masked maps: 
K and Ka, which is expected and consistent with the fact that even these 
masked maps present some level of foreground contamination away from the 
masked region (see Fig.~\ref{With_without}). The deviation for the  Q, V, 
and W masked maps are within the $95\%$ expected values from MC randomly
simulated maps.   

To have an overall quantitative assessment of the power spectra 
$S_{\ell}$ and $K_{\ell}$ calculated from  K, Ka, Q, V, W and ILC maps
we have performed a $\chi^2$ test to determine the goodness of fit 
for low $\ell$ multipole values as compared to the expected multipoles 
values from the Gaussian MC maps. In this way we have obtained numbers 
that collectively quantify the extent to which these \emph{KQ75} 
masked maps are consistent with Gaussianity. 

The results of our statistical analyses indicate that the current 
CMB temperature fluctuations ILC three and five-year masked data 
are consistent with Gaussianity, in
agreement with the WMAP team and other analyses made by using different
statistical tools~\cite{wmap1,wmap3,wmap5}.
We have demonstrated that the results of our analyses are robust by
showing that the $S-$map and $K-$map do not significantly change with
different choices of variables involved in our scheme, so long as the
statistical noise is kept under control.

The effects of different foreground-reduced algorithms as detected 
by our non-Gaussianity indicators for other WMAP three and five-year 
maps~\cite{KNC,PPG,OT} is under a careful investigation and signs of
non-Gaussianity seems to be present in the maps, which may have
a non-cosmological origin as, for example, residual foregrounds, 
artifacts of the cleaning algorithm or simply a statistical fluke.

Finally, we emphasize that the robustness of our scheme with respect
to all considered parameters along with the detection on non-Gaussianity
in the single frequency (foreground unremoved) maps seem to indicate
that our indicators are well-suited to reliably map deviation from
Gaussianity at large angular scales in the CMB data, besides being
complementary to the other approaches in the literature.

\begin{acknowledgments}
This work is supported by Conselho Nacional de Desenvolvimento Cient\'{i}fico e
Tecnol\'ogico (CNPq) -- Brasil, under grant No.\ 472436/2007-4.
M.J.R. and A.B. thank CNPq and PCI-CBPF/CNPq for the grants under which this work
was carried out. We thank an anonymous referee for her/his  constructive
and careful reports that has led to the substantial improvement of the article.
We are also grateful to A.F.F. Teixeira for reading the manuscript
and indicating the omissions and misprints.
We acknowledge use of the Legacy Archive for Microwave Background Data
Analysis (LAMBDA). Some of the results in this paper were derived using the
HEALPix package~\cite{Gorski-et-al-2005}.
\end{acknowledgments}



\begin{thebibliography}{999}
%
\bibitem{Inflation-1st-refs} A.A. Starobinsky, JETP Lett.  \textbf{30}, 682 (1979)
[Pisma Zh. Eksp. Teor. Fiz. 30, 719 (1979)];
A.A. Starobinsky, Phys. Lett. B \textbf{117}, 175 (1982);
D. Kazanas, Astrophys. J. \textbf{241}, L59 (1980);
A.H. Guth, Phys. Rev. D \textbf{23}, 347 (1981);
K. Sato, Mon. Not. Roy. Astron. Soc. \textbf{195}, 467 (1981);
A.D. Linde, Phys. Lett. B \textbf{108}, 389 (1982);
A. Albrecht and P.J. Steinhardt, Phys. Rev. Lett. \textbf{48}, 1220 (1982).

\bibitem{Inflation-reviews} B.A. Bassett, S. Tsujikawa, and D. Wands,
Rev. Mod. Phys. \textbf{78}, 537 (2006);
A. Linde, Lect. Notes Phys. \textbf{738}, 1 (2008).

\bibitem{Gauss_Single-field}
V. Acquaviva, N. Bartolo, S. Matarrese, and A. Riotto
Nucl. Phys. B \textbf{667}, 119 (2003);
J. Maldacena, JHEP 0305 (2003) 013;
M. Liguori, F.K. Hansen, E. Komatsu, S. Matarrese, and A. Riotto
Phys. Rev. D \textbf{73}, 043505 (2006).

\bibitem{Non-standard-models}
N. Arkani-Hamed, P. Creminelli, S. Mukohyama, and M. Zaldarriaga,
JCAP \textbf{0404}, 001 (2004); 
N. Bartolo, S. Matarrese, and A. Riotto, Phys. Rev. D \textbf{69}, 043503 (2004);
D.H. Lyth and Y. Rodriguez, Phys. Rev. D \textbf{71}, 123508 (2005);
G.I. Rigopoulos, E.P.S. Shellard, and B.J.W. van Tent, Phys. Rev. D \textbf{73},
083522 (2006);  
L.E. Allen, S. Gupta, and D. Wands, JCAP \textbf{0601}, 006 (2006);
X. Chen, Phys. Rev. D \textbf{72}, 123518 (2005);
N. Barnaby and J.M. Cline, Phys. Rev. D \textbf{73}, 106012 (2006);
N. Barnaby and J.M. Cline, Phys. Rev. D \textbf{75}, 086004 (2007);
N. Barnaby and J.M. Cline, 
JCAP \textbf{0707}, 017 (2007);
M. Sasaki, J. Valiviita, and D. Wands, Phys. Rev. D \textbf{74}, 103003 (2006);
X. Chen, M.-X. Huang, S. Kachru, and G. Shiu, JCAP \textbf{0701}, 002 (2007);
X. Chen, R. Easther, and E.A. Lim, JCAP \textbf{0706}, 023 (2007);
T. Battefeld and R. Easther, JCAP \textbf{0703}, 020 (2007);
H. Assadullahi, J. Valiviita, and D. Wands, Phys. Rev. D \textbf{76}, 103003 (2007);
D. Battefeld and T. Battefeld, JCAP \textbf{0705}, 012 (2007);
R. Bean, S.E. Shandera, S.H. Henry~Tye, and J. Xu, JCAP \textbf{0705}, 004 (2007);
M. Kawasaki, K. Nakayama, T. Sekiguchi, T. Suyama, and F. Takahashi,
JCAP \textbf{0811}, 019 (2008);
%
JCAP \textbf{0901}, 042 (2009); 
%
M. Kawasaki, K. Nakayama, and F. Takahashi,
JCAP \textbf{0901}, 026 (2009).

\bibitem{Bartolo2004}
N. Bartolo, E. Komatsu, S. Matarrese, and A. Riotto, Phys.
Rept. {\bf 402}, 103 (2004). 

\bibitem{wmap1}
E. Komatsu et al., Astrophys. J. Suppl. \textbf{148}, 119 (2003).

\bibitem{wmap3}
D.N. Spergel et al., Astrophys. J. Suppl. \textbf{170}, 377 (2007).

\bibitem{wmap5}
E. Komatsu et al., 
Astrophys. J. Suppl. \textbf{180}, 330 (2009).

\bibitem{Gauss-consist}
H.K. Eriksen, A.J. Banday, K.M. G\'orski, and P.B. Lilje,
\apj \textbf{622}, 58 (2005);
%
P.D. Naselsky, L.-Y. Chiang, P. Olesen, and O.V. Verkhodanov,
\apj \textbf{615}, 45 (2004);
%
P. Coles, P. Dineen, J. Earl, and D. Wright,
Mon. Not. R. Astron. Soc. \textbf{350}, 989 (2004);
%
P.D. Naselsky, O.V. Verkhodanov, L.-Y. Chiang, and I. Novikov,
Int. J. Mod. Phys. D \textbf{14}, 1273 (2005);
%
P. Naselsky, L.-Y. Chiang, P. Olesen, and I. Novikov,
Phys. Rev. D \textbf{72}, 063512 (2005);
%
K. Land and J. Magueijo, Mon. Not. R. Astron. Soc. \textbf{362}, L16 (2005);
%
P. Cabella, M. Liguori, F.K. Hansen, D. Marinucci, S. Matarrese, L. Moscardini,
and N. Vittorio, Mon. Not. R. Astron. Soc. \textbf{358}, 684 (2005);
%
M. Liguori, F. K. Hansen, E. Komatsu, S. Matarrese, and A. Riotto,
Phys. Rev. D \textbf{73}, 043505 (2006);
%
P. Cabella, F.K. Hansen, M. Liguori, D. Marinucci, S. Matarrese, L. Moscardini,
and N. Vittorio, Mon. Not. R. Astron. Soc. \textbf{369}, 819 (2006);
%
C. Hikage, T. Matsubara, P. Coles, M. Liguori, F.K. Hansen, and S. Matarrese,
Mon. Not. R. Astron. Soc. \textbf{389}, 1439 (2008);
B. Lew, JCAP \textbf{0808}, 017 (2008).

\bibitem{Chiang-et-al2003} L.-Y. Chiang, P.D. Naselsky, O.V. Verkhodanov, and M.J. Way,
Astrophys. J. \textbf{590}, L65 (2003).

\bibitem{Naselsky-et-al2005} P.D. Naselsky, L.-Y. Chiang, I.D. Novikov, and
O.V. Verkhodanov, Int. J. Mod. Phys. D \textbf{14}, 1273 (2005).

\bibitem{Eriksen-et-al2004} H.K Eriksen, F.K. Hansen, A.J. Banday, K.M. G\'orski,
and P.B. Lilje, Astrophys. J. \textbf{605}, 14 (2004).

\bibitem{Some_non-Gauss-refs}
%
P. Vielva, E. Mart\'{\i}nez-Gonz\'alez, R.B. Barreiro, J.L. Sanz, and L. Cay\'on,
Astrophys.  J. \textbf{609}, 22 (2004); 
%
M. Cruz, E. Mart\'{\i}nez-Gonz\'alez, P. Vielva, and L. Cay\'on,
Mon. Not. R. Astron. Soc. \textbf{356}, 29 (2005); 
%
M. Cruz, L. Cay\'on, E. Mart\'{\i}nez-Gonz\'alez, P. Vielva, and J. Jin,
Astrophys.  J. \textbf{655}, 11 (2007); 
%
L. Cay\'on, J. Jin, and A. Treaster,
Mon. Not. R. Astron. Soc. \textbf{362}, 826 (2005); 
%
Lung-Y Chiang, P. D. Naselsky,
Int. J. Mod. Phys. D \textbf{15},  1283 (2006);  
%
J.D. McEwen, M.P. Hobson, A.N. Lasenby, and D.J. Mortlock,
Mon. Not. R. Astron. Soc. \textbf{359}, 1583 (2005);
%
J.D. McEwen, M.P. Hobson, A.N. Lasenby, and D.J. Mortlock,
Mon. Not. R. Astron. Soc. \textbf{371}, L50 (2006);   
%
J.D. McEwen, M.P. Hobson, A.N. Lasenby, and D.J. Mortlock,
Mon. Not. R. Astron. Soc. \textbf{388}, 659 (2008);
%
A. Bernui, C. Tsallis, and T. Villela, Europhys. Lett. \textbf{78}, 19001 (2007);
%
A. Bernui, C. Tsallis, and T. Villela,
Phys. Lett. A \textbf{356}, 426 (2006);  
%
%
L.-Y. Chiang, P.D. Naselsky, and P. Coles,
Astrophys. J. \textbf{664}, 8 (2007);   
%
C.-G. Park,
Mon. Not. R. Astron. Soc. \textbf{349}, 313 (2004); 
%
D.L. Larson and B.D. Wandelt,
Astrophys. J. \textbf{613}, L85 (2004); 
%
H.K. Eriksen, D.I. Novikov, P.B. Lilje, A.J. Banday, and K.M. G\'orski,
Astrophys. J. \textbf{612}, 64 (2004); 
%
C.J. Copi, D. Huterer, and G.D. Starkman,    
Phys. Rev. D \textbf{70}, 043515 (2004);  
%
C.J. Copi, D. Huterer, D.J. Schwarz, and G.D. Starkman;
Mon. Not. R. Astron. Soc. \textbf{367}, 79 (2006); 
%
T.R. Jaffe, A.J. Banday, H.K. Eriksen, K.M. G{\'o}rski, and F.K. Hansen,
Astrophys.  J. \textbf{629}, L1 (2005); 
%
M. Cruz, M. Tucci, E. Mart{\'{\i}}nez-Gonz{\'a}lez, and P. Vielva,
Mon. Not. R. Astron. Soc. \textbf{369}, 57 (2006);    
%
M. Cruz, N. Turok, P. Vielva, E. Mart{\'{\i}}nez-Gonz{\'a}lez, and M. Hobson,
Science \textbf{318}, 1612 (2007);   
%
K. Land and J. Magueijo,
Mon. Not. R. Astron. Soc. \textbf{357}, 994 ( 2005); 
%
F.K. Hansen, P. Cabella, D. Marinucci, and N. Vittorio,
Astrophys. J. \textbf{607}, L67 (2004);   
%
P. Mukherjee and Y. Wang, Astrophys. J. \textbf{613}, 51 (2004);
%
D. Pietrobon, P. Cabella, A. Balbi, G. de Gasperis, and N. Vittorio,
\texttt{arXiv:0812.2478 [astro-ph]};
%
P. Vielva and J.L. Sanz, \texttt{arXiv:0812.1756 [astro-ph]};
%
P.K. Samal, R. Saha, P. Jain, and J.P. Ralston,
\texttt{arXiv:0811.1639 [astro-ph]}.

\bibitem{Non-Gauss_related}
E. Mart\'{\i}nez-Gonz\'alez, \texttt{arXiv:0805.4157 [astro-ph]};
%
Y. Wiaux, P. Vielva, E. Mart\'{\i}nez-Gonz\'alez, and P. Vandergheynst,
\prl \textbf{96}, 151303 (2006);
%
C.J. Copi, D. Huterer, D.J. Schwarz, and G.D. Starkman, Phys. Rev. D \textbf{75}, 023507 (2007);
%
L.R. Abramo, A. Bernui, I.S. Ferreira, T. Villela, and C.A. Wuensche, Phys. Rev. D \textbf{74},
063506 (2006);
%
D. Huterer, New Astronomy Reviews \textbf{50}, 868 (2006);
%
P. Vielva, Y. Wiaux, E. Mart\'{\i}nez-Gonz\'alez, and P. Vandergheynst,
New Astron. Rev. \textbf{50}, 880 (2006);
%
P. Vielva, Y. Wiaux, E. Mart\'{\i}nez-Gonz\'alez, and P. Vandergheynst,
Mon. Not. R. Astron. Soc., \textbf{381}, 932 (2007);
%
H.K. Eriksen, A.J. Banday, K.M. G\'orski, F.K. Hansen, and P.B. Lilje,
\apj \textbf{660}, L81 (2007);
%
K. Land and J. Magueijo, Mon. Not. R. Astron. Soc. \textbf{378}, 153 (2007);
%
A. Bernui, B. Mota, M.J. Rebou\c{c}as, and R. Tavakol,
Astron. \& Astrophys. \textbf{464}, 479 (2007);
%
A. Bernui, B. Mota, M.J. Rebou\c{c}as, and R. Tavakol,
Int. J. Mod. Phys. D  \textbf{16}, 411 (2007);
T. Kahniashvili, G. Lavrelashvili, and B. Ratra, Phys. Rev. D \textbf{78}, 063012 (2008);
B. Lew, JCAP \textbf{0809}, 023 (2008);
%
A. Bernui, Phys. Rev. D \textbf{78}, 063531 (2008);
F.K. Hansen, A.J. Banday, and K.M. G\'orski, Mon. Not. R. Astron. Soc. \textbf{354}, 641 (2004);
%
F.K. Hansen, P. Cabella, D. Marinucci, and N. Vittorio,
\apj \textbf{607}, L67 (2004);
%
P. Bielewicz, K.M. G\'orski, and A.J. Banday,
Mon. Not. R. Astron. Soc. \textbf{355}, 1283 (2004);
%
K. Land and J. Magueijo, \prl \textbf{95}, 071301 (2005);
%
%
A. Bernui, T. Villela, C.A. Wuensche, R. Leonardi, and I. Ferreira,
Astron. \& Astrophys. \textbf{454}, 409 (2006);
%
M. Tegmark, A. de Oliveira-Costa, and A.J.S. Hamilton,
Phys. Rev. D \textbf{68}, 123523 (2003);
%
A. de Oliveira-Costa, M. Tegmark, M. Zaldarriaga, and A. Hamilton,
Phys. Rev. D \textbf{69}, 063516 (2004);
%
J.R. Weeks, astro-ph/0412231;
%
P. Bielewicz, H.K. Eriksen, A.J. Banday, K.M. G\'orski, and P.B. Lilje,
\apj \textbf{635}, 750 (2005).


\bibitem{ILC-WMAP-refs} G. Hinshaw \emph{et al.}, Astrophys. J. Suppl.
\textbf{170}, 288 (2007); 
G. Hinshaw \emph{et al.}, 
Astrophys. J. Suppl. \textbf{180}, 225 (2009).

\bibitem{Gorski-et-al-2005} K.M. G\'orski, E. Hivon, A.J. Banday, B.D. Wandelt,
F.K. Hansen, M. Reinecke, and M. Bartelman, Astrophys. J. \textbf{622}, 759 (2005).

\bibitem{wmap5-Gold-et-al} B. Gold \emph{et al.}, 
Astrophys. J. Suppl. \textbf{180}, 265 (2009).

\bibitem{wmap3-Spergel-et-al} D.N. Spergel, \emph{et al.},
Astrophys. J. Suppl. \textbf{170}, 377 (2007).  

\bibitem{KNC} J. Kim, P. Naselsky, and P.R. Christensen,
Phys. Rev. D \textbf{77}, 103002 (2008).

\bibitem{PPG} C.-G. Park, C. Park, and J.R. Gott III,
Astrophys. J. \textbf{660}, 959 (2007).

\bibitem{OT} A. de Oliveira-Costa and M. Tegmark, Phys. Rev. D \textbf{74}, 023005 (2006).
astro-ph/0603369
CMB multipole measurements in the presence of foregrounds




\end{thebibliography}
\end{document}